\documentclass[12pt]{spieman}  
\usepackage[]{graphicx}
\usepackage{setspace}
\usepackage{epstopdf}
\usepackage{tocloft}
\usepackage{amssymb}
\usepackage{amsmath} 
\usepackage{deluxetable}
\usepackage{subcaption}

\title{Strategies for Detecting the Missing Hot Baryons in the Universe} 

\author{Joel N. Bregman,\supscr{a} Guilherme C. Alves,\supscr{a}, Matthew J. Miller\supscr{a}, and Edmund Hodges-Kluck\supscr{a}}
\affiliation{\supscrsm{a}Department of Astronomy, University of Michigan, Ann Arbor, MI  48109}

\cftpagenumbersoff{figure}
\cftpagenumbersoff{table} 
\begin{document} 
\maketitle 

\begin{abstract}

About 30-50\% of the baryons in the local Universe are unaccounted for and are likely in a hot phase, $10^{5.5}-10^8$ K.  A hot halo ($10^{6.3}$ K) is detected around the Milky Way through the O VII and O VIII resonance absorption and emission lines in the soft X-ray band.  Current instruments are not sensitive enough to detect this gas in absorption around other galaxies and galaxy groups, the two most likely sites.  We show that resonant line absorption by this hot gas can be detected with current technology, with a collecting area exceeding about 300 cm$^2$ and a resolution R $>$ 2000.  For a few notional X-ray telescope configurations that could be constructed as Explorer or Probe missions, we calculate the differential number of O VII and O VIII absorbers as a function of equivalent width through redshift space, $dN/dz$.  The hot halos of individual external galaxies produce absorption that should be detectable out to about their virial radii.  For the Milky Way, one can determine the radial distribution of density, temperature, and metallicity, after making optical depth corrections.  Spectroscopic observations can determine the rotation of a hot gaseous halo.

\end{abstract}


{\noindent \footnotesize{\bf Address all correspondence to}: Joel N. Bregman, Department of Astronomy, University of Michigan, 1085 South University Ave., Ann Arbor, MI  48109; Tel: +1 734-764-3440; Fax: +1 734-764-6317; E-mail:  \linkable{jbregman@umich.edu} }

\begin{spacing}{1}   

\section{Introduction}
\label{sect:intro}  

Despite the prominence of galaxies, they only account for about 5\% of the baryons in the Universe, the remainder still in the gaseous form \cite{fuku98}.  Much of this gaseous component has collapsed into filaments where it is shocked and the higher overdensity regions have led to the formation of galaxies, galaxy groups, and galaxy clusters of today  \cite{mo2010, krav12, schaye15}.  Feedback from stellar evolution, stellar mass loss and supernovae, and from AGNs have heated and polluted the gaseous medium, expelling much of the gas from galaxies.  The history of these processes is encoded in the properties of this gaseous component, motivating the enormous amount of activity over the past several decades.

Absorption line studies have established the properties of neutral hydrogen, through Lyman series absorption \cite{wolfe05}, and of the metals, from resonance lines of common elements in several ionization states \cite{stocke13, werk14}.  Most of the mass lies in the higher column density systems, but the hosts of the neutral hydrogen are somewhat different than that of the metals.  At the higher column densities, the absorption is related to galaxies, although the spatial distribution depends on the ionization state.  From absorption studies, a census at low redshift shows that most of the known baryons are in gaseous form, with the largest component associated with gas in the $10^4 - 10^{5.5}$ K
range.  However, about 30-50\% of the baryons are unaccounted for by these studies \cite{shull12}, suggesting that a considerable amount of gas lies in a temperature range that does not produce UV absorption lines.

A related census is that of the cosmic density of metals, which can be estimated from the cosmic star formation rate and the metals lost during normal stellar evolution \cite{pettini06, shull14}.  This comparison shows that only 10\% of the metals are accounted for through the stars and UV quasar absorption line studies and this is the case both in the local universe and at z = 2-3.  Evidently, most of the metals and about half of the baryons are yet to be detected.

The physical state and location of the baryons is suggested by simulations of structure formation, which indicates that infall onto massive galaxies and larger virialized structures (e.g., galaxy clusters) will lead to extended hot gas distributions with temperatures of $10^{5.5}-10^8$ K \cite{hayward14}.  In addition, the collapse of unvirialized filaments also produces hot gas through the density wave shocks that develop ($10^5 - 10^6$ K).  For temperatures below $10^{5.5}$ K, high ionization UV lines (e.g., O VI, C IV) are powerful diagnostics, but at temperatures above about $10^{5.5}$ K, the dominant ions have their ground state resonance transitions in the X-ray band \cite{bdsmith11, cen12}.

The UV absorption line studies show that O VI is commonly observed near galaxies, but there may also be a major contribution from the medium of galaxy groups \cite{stocke14}.  The characteristic temperature of O VI, $10^{5.3}$ K, is an order of magnitude below the virial temperature of these galaxies and galaxy groups, so a hotter medium should be present, detectable at X-ray wavelengths from ions such as O VII and O VIII.  This hotter medium is detected through X-ray emission around massive galaxies.  Studies of relatively isolated early and late type galaxies show that there are hot gaseous halos extending to at least 50 kpc (0.1-0.2 R$_{200}$), beyond which, diffuse X-ray emission becomes increasingly difficult to detect \cite{anderson11, humph12, dai12, bogdan13, walker14}.  If this gas extends beyond 50 kpc, a more sensitive measure is needed, which can be provided by X-ray absorption techniques.

There is good evidence that hot gas extends at least to the virial radius and probably beyond.  The evidence comes from the S-Z studies carried out by the Planck Observatory, where they stack a large number of galaxies of different masses, obtaining a significant signal for galaxies with M* $ > 10^{11}$ M$_{\odot}$ \cite{planckXI13, greco14}.  For these luminous galaxies, the signal implies that most of the baryons are near the virial temperature of the system.  If this result can be smoothly extrapolated to less massive and more common galaxies, it implies that they too are primarily hot gaseous systems.  We note that this result appears to be in conflict with the UV absorption line results of Ref.\citenum{werk14}, who argue that most of the baryons are in a much cooler stage, closer to $10^4$ K.  This tension between two approaches needs to be resolved.

For the column densities that are expected around massive galaxies, galaxy groups, and possibly other parts of the intergalactic medium (outskirts of galaxy clusters; denser filaments in the cosmic web), detections in the X-ray resonance lines are possible with improvements in instrumentation.  Current instruments, such as the RGS on \textit{XMM-Newton} and the LETG on the \textit{Chandra Observatory}, detect X-ray absorption lines from the Milky Way halo, validating the approach of using X-ray resonance lines \cite{nica02,fang02, rasm03,yao05,mckernan04,wang05,williams05,miller13}.  Beyond the Milky Way, the reliability of X-ray absorption lines is poor \cite{breg07}, as there may have been some confusion between absorption features and galactic absorption by previously unknown inner shell lines \cite{nica14}.  Extragalactic absorption was not predicted to be detectable with current instrumentation (\textsection 4), so this is not a blow to the approach.

Extragalactic X-ray absorption line studies will become a burgeoning field with an improvement in detection efficiency of an order of magnitude \cite{yao12}.  The improvement can be realized in both spectral resolution and collecting area.  The spectral resolution of the \textit{XMM-Newton} RGS and \textit{Chandra} LETG is about 360-440 for the O VII and O VIII lines at z = 0, and with collecting areas is 43 cm$^2$ and 3 cm$^2$ (see Table 1 for a listing of related instrumental properties along with those considered here).  Within the typical cost of previous \textit{NASA} Explorer and Probe class missions, it is possible to obtain a spectral resolution of about 3000 \cite{smith14}, which is about the Doppler width of the lines, and a collecting area that approaches 1000 cm$^2$.  Together, these provide more than an order of magnitude improvement in line detectability.  Several concepts along these lines have either been proposed or are in development, so here we outline a plan of how to best use these resources to construct an instrument at the Explorer or Probe class level to investigate absorbing hot gas in galaxy halos, galaxy groups or clusters, and in the cosmic web.

In this paper we consider several issues surrounding a potential future mission, extending the discussion by Ref.\citenum{yao12} in a variety of ways. We identify the likely targets that can be used (\textsection 2) and calculate the observing time required for useful detections.  Observing strategies are discussed for the testing of cosmological predictions, to detect the halos of galaxies, and for studies of the Milky Way and M 31 (\textsection 3).  Further discussion and conclusions are given in \textsection 4.

\section{Constructing the Observing List} 

A central requirement of a successful program is that enough absorption line systems can be detected to make scientific progress.  One consideration is  that enough detections be obtained such that one begins to have an understanding of the space density of absorption systems as a function of column density ($dN/dz$).  This column density distribution determines the metallicity of these highly ionized metals as a function of cosmic volume, which is a critical measurement given that 90\% of metals in the low redshift universe are ``missing".  The column density distributions are predicted from simulations \cite{cenfang06, branchini09, cen12} and provide a starting point from which to plan a program.

\subsection{The Relevant Absorption Lines}

A basic consideration is that one has the sensitivity to probe the anticipated sites of hot gas absorption.  The expected sites are galaxies, galaxy groups, and galaxy clusters, which have characteristic virial temperatures of $10^6 - 10^7$, $10^{6.5} - 10^{7.5}$, and $10^{7.5} - 10^{8.2}$ respectively.  In this energy range, we compile a list of the most prominent resonance lines of common elements, as shown in Table 2.  The temperature range of the ion roughly increases with the line energy, which might suggest that one can probe gas through temperatures of $10^8$ K, using highly ionized Fe.  However, the fractional equivalent width of a line decreases linearly with the wavelength, so increasingly hotter gas would demand an equivalent increase in the S/N of the continuum to obtain a detection.  Another consideration is the relative abundance, where the lighter and more common elements (O, C, N) generally have their resonance lines at lower energies than the heavier species (Ne, Mg, Fe, Si).  This leads to O VII and O VIII being the most easily detected resonance lines, followed by those of C, N, Ne, and Fe.  For these reasons, most of the sensitivity calculations and discussion will focus on the O VII and O VIII resonance lines.

The strongest resonance lines are associated with gas in the range $10^6-10^7$ K, typical of galaxy halos and the less massive galaxy groups.  The Milky Way is known to produce X-ray absorption lines \cite{miller13}, and many galaxies show absorption from O VI \cite{werk14}, the ion adjacent to O VII.
The common lower-mass galaxy groups (10$^{13}-10^{14}$ M$_{\odot}$) may be good sites for absorption, as the characteristic temperature can be below $5 \times 10^6$ K \cite{osullivan14}, and in which the lower temperature ion, O VI, may be quite plentiful \cite{stocke14}. 
These are often the sites of X-ray absorption in the simulations of Ref.\citenum{cenfang06,branchini09,cen12}.
In contrast, the region within $R_{500}$ in clusters and rich groups have temperatures well above $10^7$ K, making them a poor place to search for X-ray absorption lines.  However, if the temperature decreases significantly at or beyond $R_{200}$ in galaxy clusters \cite{urban14}, this will be a possible absorption site.  

\subsection{Energy Range Constraints}

\begin{figure}
	\begin{center}
		\begin{tabular}{c}
			\includegraphics[height=10.cm]{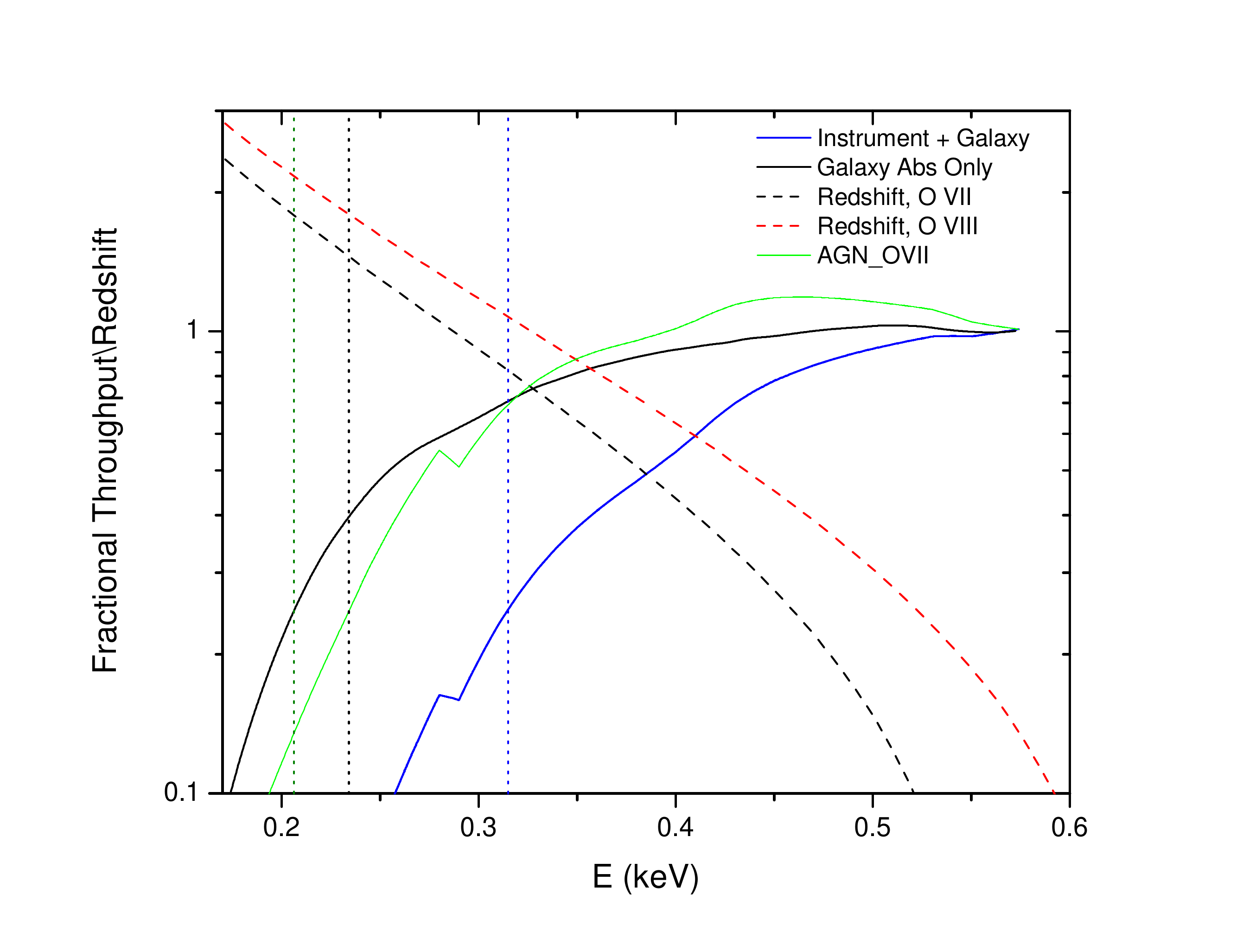}
		\end{tabular}
	\end{center}
	\caption{
		The throughput of photons relative to the rest energy 
		of the O\ VII resonance line (0.574 keV) as a function of decreasing energy,
		which corresponds to increasing redshift. \ The solid black line shows the
		throughput change due to Galactic absorption with a neutral hydrogen column
		of $2\times 10^{20}$ cm$^{-2}$, while the blue solid line combines the
		response of the Arcus spectrograph \cite{smith14} with this Galactic absorption. The green line 
		is the throughput when including a power-law continuum typical of an AGN.
		The dashed black and red lines show the increase of the redshift of the resonance 
		lines of O VII and O VIII with decreasing energy, while the three vertical 
		dotted lines show the 25\% transmission for Galactic absorption (left line) 
		plus the effective area decrease (right line), an including a typical AGN 
		spectrum (middle line).
	}
	\label{fig:Throughput}
\end{figure}

The detectability of the O\ VII and O\ VIII\ resonance lines generally
decreases with increasing redshift due to two effects: \ Galactic
absorption; and a decline in instrumental sensitivity. \ For a typical
sightline out of the plane of the Milky Way, the column density of neutral
gas produces an increasing optical depth with decreasing energy, largely due to
He in the energy range of interest. \ The
column densities generally lie in the range of $1-10\times 10^{20}$ cm$^{-2}$, 
so we illustrate this point by taking a column of $2\times 10^{20}$ cm$%
^{-2}$, a moderately favorable line of sight. \ At this column, and for an
instrument with a response that is independent of energy, the throughput,
relative to the rest energy of O\ VII, falls by a factor of four by an
energy of 0.207 keV, which is a loss in S/N of a factor of two (Figure~\ref{fig:Throughput}). \ The locus for O\ VIII is very similar, so we use the O\ VII
calculation. At 0.207 keV, the redshifts of the O\ VII and O\ VIII resonance
lines are 1.76 and 2.14.

When one takes into account the change in effective area of a typical
mirror-instrument combination, the situation worsens. \ Here we use the
Arcus effective area \cite{smith14}, although we would obtain about the same
result when using the Athena microcalorimeter (X-IFU) response \cite{denHerder12}. \
The effective area decreases with decreasing energy, and when combined with
Galactic absorption, leads to a loss of throughput by a factor of four at
0.315 keV, which corresponds to redshifts of 0.83 and 1.08 for O\ VII and O
VIII. \ An order of magnitude loss in throughput occurs at an energy of 
0.251 keV, with corresponding redshifts of 1.22 and 1.53 for O\ VII and O\ VIII. 

Some of the decline in throughput at low energies is mitigated if one has a 
continuum that rises toward low energies, such as an AGN continuum without 
significant intrinsic absorption.  For an AGN with no internal absorption and
a power-law of index 1.7, the throughput falls to one-quarter of its value
at 0.574 keV at an energy of 0.237 keV (Figure~\ref{fig:Throughput}).  
This corresponds to a  redshift of 1.43 for O VII and 1.78 for O VIII.  
For these reasons, and to be conservative, we do not consider
redshifts above 1.0 in the calculations below. 
We note that our calculations suggest that redshift space studies can be
improved if the blocking filters can be designed to have higher throughput
at energies below about 0.4 keV, as it is these filters that account for most 
of the decline in the effective area.

\subsection{Selecting Potential Targets from Archival X-Ray Data}

The study of the warm-hot intergalactic medium (WHIM) requires that targets have a bright continuum in the X-ray range against which absorption lines can be detected. These targets were found with data from various different X-ray catalogs in the 0.5-2 keV energy band, such as bright sources from $\textit{ROSAT}$ and $\textit{XMM-Newton}$.  The $\textit{ROSAT}$ Bright Source Catalog (BSC)\cite{voges99} was cross-correlated with the AGN catalog of Ref.\citenum{veron10}, with corrections \cite{flesch13}, in order to obtain a list the X-ray brightest AGNs.  Data from $\textit{Chandra}$, $\textit{Swift}$, and $\textit{Suzaku}$ were also examined to obtain the most complete list, as we are also interested in targets that have historically flared but have a low median flux. Low redshift targets were generally excluded from the list, as they do not offer a significant amount of redshift space for searches.
Some bright low redshift objects are included, as they are useful for Galactic studies.  However, we caution that some AGN have their own blueshifted absorption lines that might be confused with intergalactic absorption lines.

This approach led to a list of about 200 sources, and to make a smaller more strategic list, we compiled the published fluxes along with their variations.  As we are primarily interested in the absorption below 1 keV, we used the commonly published 0.5-2 keV flux and adjusted the fluxes from the various catalogs, which are given in a variety of different energy bandpasses.    

The three primary catalogues from which fluxes were obtained are: the $\textit{ROSAT}$ All-Sky Survey: Bright Sources \cite{voges99}; the $\textit{ROSAT}$ Complete Results Archive Sources for the PSPC; and the $\textit{XMM-Newton}$ Serendipitous Source Catalog \cite{watson09} (3XMM DR4 Version). The data from $\textit{ROSAT}$ is given in counts/sec, so we applied a spectral model, which was a power-law spectrum with photon index $\Gamma = 2$ and with Galactic absorption, where the Galactic HI colums are given by the Leiden/Argentine/Bonn (LAB) Survey of Galactic HI \cite{kalberla05}, since this value could change the resulting converted flux in erg cm$^{-2}$ s$^{-1}$. For 1 cnt/sec, and where $N_H$ is the Galactic HI column, the conversion can be parameterized as
\begin{equation}
F_X (0.5-2 keV) = 1.135 \times 10^{-11} - 7.561 \times 10^{-12} \exp(-N_H/2.89 \times 10^{20}) .
\end{equation}

The $\textit{XMM-Newton}$ fluxes are given in nine different energy bands (E1 to E9). To have an equivalent value for the 0.5-2.0 keV band, one combines the E2 and E3 fluxes, which corresponds to 0.5-1.0 keV and 1.0-2.0 keV, respectively. 
Both the $\textit{ROSAT}$ Complete Results Archive Sources for the PSPC and the $\textit{XMM-Newton}$ Serendipitous Source Catalog had multiple measurements of flux F$_X$, so the final value for these fluxes was the average of all the measurements. 
We quantify the variation about the mean by calculating the standard deviation of the objects in Table 3 as a percentage value.  Typically, an individual object has a variation of about 30\% for a single observation relative to the mean, but some objects exhibit much larger variation, such as MRK 421, whose flux varies more than 75\%.

Subsequent to data collection, we ranked the candidates according to a figure of merit, which is defined as approximately the number of absorption systems detected per unit observing time. The figure of merit is dependent on the amount of redshift space that can be probed and the number of absorption systems predicted per unit redshift as a function of the continuum S/N. We adopt $Z_{AGN}$ as the redshift space being adopted, although there are some small corrections, which would hardly have an effect on the ranking. These corrections are mostly due a small redshift region around the AGN ($\sim$5000 km/s), where the material surrounding the AGN may cause absorption that is unrelated to the IGM, such as AGN winds or accretion. Another cause can be the small region of velocity space near $z = 0$ ($\sim$400 km/s) caused by Milky Way absorption features, also unrelated to the IGM.

We define a figure of merit under the assumption that all spectra have the same value of S/N (equivalently, a single limiting equivalent width; henceforth EW). Thus the observing time is inversely proportional to the flux density of the background AGN in the energy range of interest. The limiting EW is related to the total number of absorption lines per unit redshift, $dN/dz(>EW)$, by the simulations of Ref.\citenum{cenfang06}, and for EW in the range 1-14 m\AA, can be fit with a polynomial, expressed as (EW given in m\AA)
	\begin{equation}
	\log{\frac{dN}{dz}}= 1.211 -0.598 \cdot \log(EW) -1.36 \cdot (\log(EW))^2 + 2.62 \cdot (\log(EW))^3   -1.96 \cdot (\log(EW))^4
	\end{equation}
Then, to within a constant, we define a relative figure of merit as
	\begin{equation}
	M = z\cdot \frac{dN}{dz}\cdot F_X
	\end{equation}
where $F_X$ is the flux density in the relevant energy range.  For the flux density, we use the value in the 0.5-2 keV range. For the redshift measure, we limit the values to less than unity to avoid effects of Galactic absorption and a decrease in instrumental sensitivity (see above); we use the minimum of $z = 1$ and $z_{AGN}$. This figure of merit was based in a number of simplifying assumptions, such as that the instrumental sensitivity (spectral resolution and collecting area) is independent of wavelength and that the functional form of  $dN/dz$ is independent of redshift up to $z = 1$.  However, it is useful to define the best candidates for the purpose of detecting absorption lines in absorption lines in the IGM.

Other than merit, we can also rank all the objects regarding the simulated predictions for their exposure time. The expression for exposure time was created from \textit{XSPEC} simulations and compared to the detections of known sources with the \textit{XMM} RGS \cite{miller13}.  It is expressed as:
\begin{equation}
t_{exp} = 0.31 \left(\frac{1\times 10^{-11}}{F_x(0.5-2~\textrm{ keV})}\right) \left(\frac{3000}{R}\right) \left(\frac{1000 ~\textrm{cm}^{2}}{A}\right) \left(\frac{0.5 ~\textrm{m\AA }}{\sigma}\right)^2  \textrm{Msec},
\end{equation}
where $\sigma$ refers to the uncertainty in the value of the EW.
We calculated the exposure times, redshift space probed, and the expected number of absorbing systems in Table 4, for a telescope with a collecting area of 1000 cm$^2$, R = 3000, and for a limiting EW of 3 m\AA\ and an error in EW of 0.6 m\AA\ (a 5$\sigma$ detection).
We set a minimum exposure time of 0.1 Msec.  For the brightest sources that might need less than 0.1 Msec, this minimum exposure time leads to a higher S/N and the investigation of the number density of lower EW systems at relatively low observing cost.

After assembling a list of the best 100 objects, we searched each field with archival optical imaging to identify any relevant foreground objects (i.e., galaxies, galaxy groups, galaxy clusters) and to determine if the AGN was in a galaxy group or galaxy cluster.  The archival imaging data used was from the Palomar Digital Sky Survey (DSS) and from the Sloan Digital Sky Survey III Data Release 10 (SDSS DR10\cite{ahn14}), although we also utilized literature images and searched for $\textit{HST}$ images when relevant.  Given this imaging set, most of the identifications are for galaxies closer than z = 0.3, and there is significant incompleteness, especially when only DSS images were available, so the frequency of our identifications is a lower limit to what would be found with a dedicated search.
Some clusters and groups were already cataloged in the literature and referenced through NED studies.  

Galaxies were considered to be associated with the AGN for velocity differences of 1000 km s$^{-1}$ in the case of a group and for twice the measured velocity dispersion of a galaxy cluster, although associations with galaxy clusters was uncommon.
Determining associations is easily accomplished when we have redshifts available from spectra, such as from SDSS or from studies in the literature.
However, in some cases there are only photo-z redshifts, mostly from the SDSS KD-tree and RF methods, so based on their uncertainties and whether the range of possible redshifts overlapped with the AGN redshift, we assigned it to the AGN group.  Potential intervening galaxies with no redshift information were not included in Table 3.
To claim an association with a potential intervening galaxy group, the background AGN had to lie within 3 Mpc of the group center.  We note that most galaxy group surveys do not probe the very common modest groups ($<10^{14}$ $M_{\odot}$), a situation that should be improved upon with upcoming surveys this decade.

There is no well-defined X-ray impact parameter to help us define that a galaxy is sufficiently close, but in the UV, the low and medium ionization absorption lines are seen to about 150 kpc from luminous galaxies, with the higher ionization ions having a slower decline in EW with radius \cite{werk14}.  For the purposes of this work, we identify luminous galaxies not in the AGN galaxy group that are projected within 200 kpc from the line of sight as being of possible special interest, although we acknowledge that this is an arbitrary choice.  In seeking luminous galaxies, we adopted a lower stellar mass bound of $10^{10}$ M$_{\odot}$.  Stellar masses were estimated from the \textit{r} band magnitudes, which were commonly available, using a value for a M* galaxy of M(r)=-21.16 (a stellar mass of $10^{10.2}$ M$_{\odot}$).  Galaxies that were either fainter or more distant may also be of some interest and were also noted, as were galaxies within galaxy groups, half of which lie in front of the AGN.  We designate these different situations in Table 3 with the numbers:  1 for a luminous intervening galaxy projected within 200 kpc of the AGN sightline; 1.5 for a lower luminosity galaxy (down to M(r) = -18) projected within 200 kpc of the AGN or a high luminosity galaxy projected between 200-400 kpc from the AGN; 2 is for a galaxy within the same group or cluster as the AGN that is projected within 200 kpc and is a luminous system.

\subsection{The Viability of Using GRBs as Background Continuum Sources}

\begin{figure}
	\begin{center}
		\begin{tabular}{c}
			\includegraphics[height=10.cm]{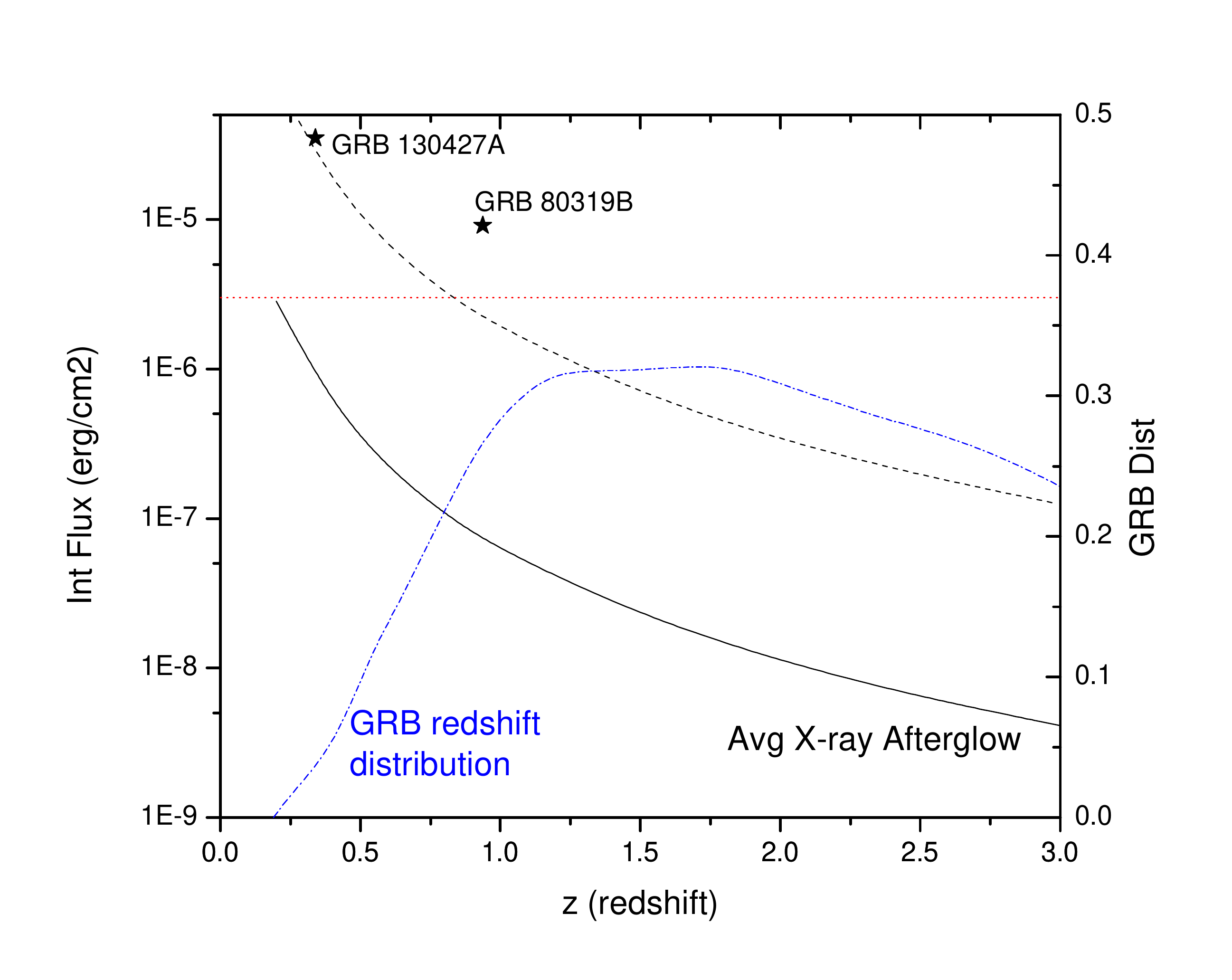}
		\end{tabular}
	\end{center}
	\caption{
		The integrated 0.5 - 2.5 keV flux of the average X-ray afterglow of long GRBs (solid black line) vs redshift. The integrated flux that is 10 times the average is the dashed black line, the dotted horizontal line is a typical integrated flux needed for a continuum against which IGM absorption can be detected 3$\times$10$^{-6}$ erg cm$^{-2}$.  The redshift distribution (blue dash-dot line and right scale) showing a minority of GRBs at low redshift. The integrated X-ray flux of two of the brightest GRBs in several decades is shown.
	}
	\label{fig:GRB}
\end{figure}

Another potential class of bright background sources are the Gamma Ray Bursters (GRB), especially the long bursts, which are seen to high redshift and can be detected for days.  Associated with GRBs are X-ray afterglows, where the average time dependent flux and the distribution of X-ray luminosities are known from \textit{Swift} observations \cite{gehrels09}.  Nearly all of the X-ray afterglows lie within an order of magnitude of the mean X-ray luminosity (the luminosity integrated over the initial 2$\times$10$^5$ sec), so we consider the output of the average GRB as well as those up to an order of magnitude above the average.  These are shown as a function of redshift (Figure~\ref{fig:GRB}), along with a typical integrated flux target that we adopt for AGN sources, 3$\times10^{-6}$ erg. 

The average GRB would need to be closer than a redshift of about 0.2, but there are very few such long GRBs, as the median redshift is about 2.2 and the distribution falls rapidly for $z < 1$ \cite{coward13}.  Even the objects an order of magnitude more luminous than the average would need to lie below z = 1 to accumulate enough photons for the required high S/N continuum.  In addition, about three-quarters of GRBs have a high amount of X-ray extinction \cite{starling13}, which removes much of the continuum in the energy range 0.25 - 0.574 keV, where one would seek to detect O VII.  There are few sufficiently luminous and unabsorbed GRBs with z $<$ 1, indicating that there would be fewer than one suitable GRB per year that can be used as background sources.  A couple of notable examples, 10-30 times brighter than the mean X-ray afterglow are GRB 130427A \cite{perley14}, the brightest GRB in 29 years (z = 0.34) and GRB 080319B \cite{racusin2008}, the “naked eye” GRB (z = 0.94) and one of the most luminous and brightest ever recorded.  As discussed below, modeling suggests that we need $\sum z \approx 10$ to characterize intergalactic absorption and this goal would take more than a decade if one used GRBs as background sources.

Spectra of GRBs may be quite revealing of another phenomena, the high absorption associated with most of the spectra \cite{starling13}.  When interpreted as absorption intrinsic to the GRB, the enormous columns, $\sim 10^{22}$ cm$^{-2}$ can create large equivalent widths that can be detected against lower S/N continua than those needed for detection of IGM lines.  The implications for such studies are beyond the scope of this paper.

\subsection{Observing Strategy} 
\label{sect:ObsStrategy}

There are a variety of strategies that can be employed for detecting intergalactic absorption lines, which depend upon the goals that are set forth and the instruments available.  If the goal is simply to accumulate the maximum amount of absorption line systems in a blind survey, and if targets never varied, one would simply observe the targets in descending order of merit, as delineated in the previous section.   However, there might be other considerations, such as choosing sightlines near objects of known interest, or determining $dN/dz$ as a function of redshift space.

\subsection{Defining the $dN/dz$ Distribution of Absorption Systems} 

\begin{figure}
	\begin{center}
		\begin{tabular}{c}
			\includegraphics[height=10.cm]{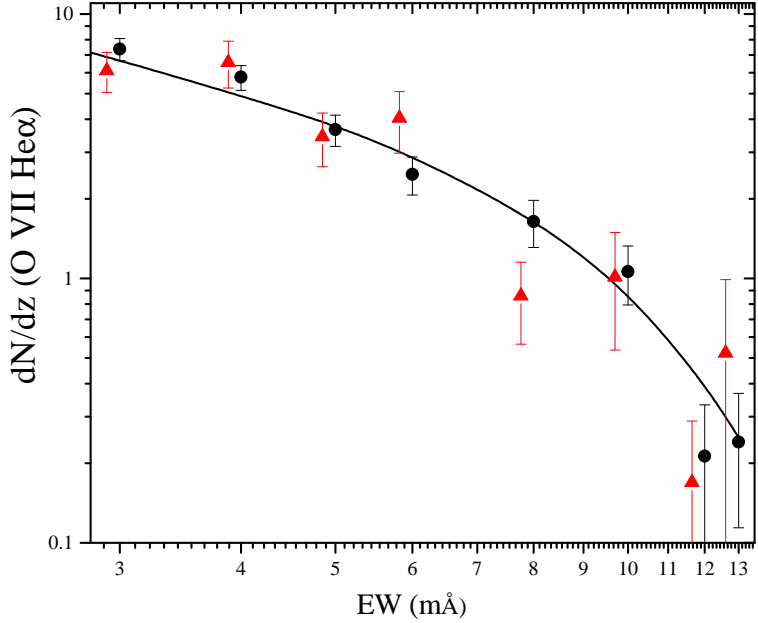}
		\end{tabular}
	\end{center}
	\caption{
		The $dN/dz$ for the O VII He$\alpha$ 21.60 \AA\ line from the cosmological simulation by Ref. \citenum{cenfang06} (solid black line), along with simulated observations of 100 absorption line systems to (black points) and for 33 absorption line systems (red triangles; offset in EW for readability) to a uniform limiting equivalent width of 3 m\AA .  The definition of the distribution at large EW could be improved for a strategy that includes observing additional sight lines to a higher limiting EW.
	}
	\label{fig:dNdz}
\end{figure}

The frequency of absorbing systems as a function of equivalent width is a broadly used function that naturally summarizes observational surveys and can be connected to theoretical simulations.  The functional form of this distribution (e.g., Ref.\citenum{cenfang06}) has a normalization, a power-law slope up to some EW and then an exponential cutoff (Figure~\ref{fig:dNdz}).  The normalization is largely set by the characteristic metallicity of the absorbing gas, the power-law slope by the large-scale structure distribution, and the cutoff by the product of the metallicity and the maximum column density in cosmic structures for the relevant temperature range.  For a limiting EW of 10 m\AA, which is in the cutoff region, Ref.\citenum{cenfang06} predict about one absorption system over a redshift search space of unity, while there are seven absorption systems if the limiting EW is 3 m\AA.  Ideally, one should define $dN/dz$ over the largest range of EW space as possible.

\begin{figure}
	\begin{center}
		\begin{tabular}{c}
			\includegraphics[height=10.cm]{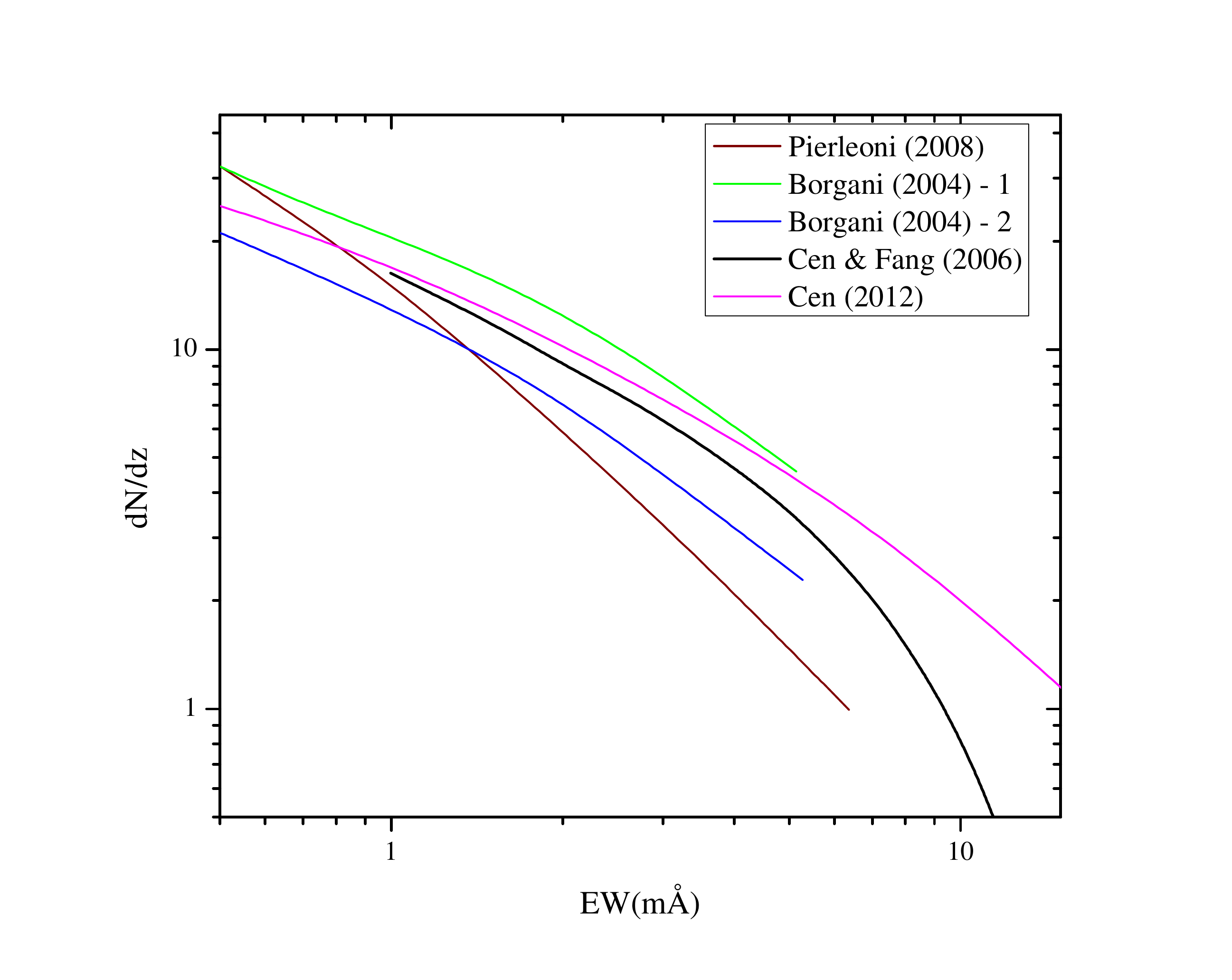}
		\end{tabular}
	\end{center}
	\caption{
		The $dN/dz$ distribution for the O VII He$\alpha$ 21.60 \AA\ absorption line for several different models, with those by Ref. \citenum{cenfang06} and Ref. \citenum{cen12} covering the broadest EW range. 
		The distributions are taken from published figures, so the endpoints depend on those figures and do not reflect the true ends of the distributions. 
		For calculation purposes, we adopt the functional form of Ref. \citenum{cenfang06}, which lies near the middle of the various models.
	}
	\label{fig:dNdzVarious}
\end{figure}

There have been other calculations for $dN/dz$, which are within about 30\% of the Ref.\citenum{cenfang06} values (Figure~\ref{fig:dNdzVarious}).  The more recent and more accurate hydrodynamic calculations by Ref.\citenum{cen12} gives a $dN/dz$ distribution that lies about 14\% above that of Ref.\citenum{cenfang06} at 3 m\AA\, but has a more significant tail at high equivalent width, lying about a factor of two higher at 10 m\AA .  
Ref. \citenum{branchini09} took simulations from other investigators and calculated $dN/dz$, where the papers chosen use an SPH approach (GADGET-2 \cite{spring05}).  
He uses two simulations from Ref.\citenum{borgani04} that use somewhat different physics and feedback and this leads to values of $dN/dz$ being about 30\% higher and lower than Ref.\citenum{cenfang06} at 3 m\AA .  
\citenum{branchini09} also considers a calculation by Ref.\citenum{pier08}, which has a very different shape than all the other distributions.  It is a steep function, lying above the other distributions at low EW and below them at high EW; at EW = 3 m\AA\, it is 50\% below the value given by Ref.\citenum{cenfang06}.  There is some concern that SPH codes predict too few absorption systems \cite{cen12}, possibly due to poor mixing of the metals.
Additional calculations for $dN/dz$ could be calculated from modern simulations, such as that of Ref.\citenum{shull12}.  In this paper, we adopt the functional form of $dN/dz$ given by Ref.\citenum{cenfang06}, which lies about in the middle of the various calculations.  This is probably a conservative choice as the more recent work of Ref.\citenum{cen12}, which reproduces the O VI data, would predict more O VII absorption lines.

The easiest aspect of $dN/dz$ to determine is the normalization, and since it is just proportional to the square root of the number of systems, the normalization is determined to 10\% with 100 absorption systems and 30\% with 10 systems (3 m\AA\  threshold).  To obtain 100 O VII absorption systems, one would have to search over an accumulated redshift space $\sum z \approx 15$ (Table 4), while 10 systems would result from a search space of $\sum z \approx 1.5$.  The turnover region is defined by strong but rare absorption line systems (EW = 10-15 m\AA).  To place constraints on it, one needs to sample as much redshift space as possible but a higher threshold is acceptable, such as 10 m\AA\  over at least $\Sigma z = 10$, which will yield about 10 sources with EW above 10 m\AA.  At the higher EW threshold, the observing time per object is about an order of magnitude less, which suggests a strategy with a deep survey to a 3 m\AA\ threshold, supplemented by a shallower survey of additional objects to a 10 m\AA\  threshold.  A more complicated multi-tiered approach might be more efficient, but as we currently have no data, it is a strategy that could be implemented once basic information about $dN/dz$ has been obtained.  In summary, the full shape of the $dN/dz$ distribution would be well-defined by about 100 absorption systems, appropriately distributed over EW space.  Fewer absorption systems will still provide vital information about $dN/dz$, but with larger uncertainties.

The $dN/dz$ associated with O VIII, when compared to that of O VII, is a temperature sensitive measure of the hot absorbing medium and provides constraints on the degree of heating during the formation of structure process.  The O VIII absorption line systems are a bit less common, but the difference is not even a factor of two at most EWs \cite{cenfang06}, so a good survey of the O VII absorption lines will provide about half the number of O VIII absorption lines.

A good observational goal is to have a search space of $\sum z = 10$ with a 3 m\AA\  threshold (5$\sigma$), as discussed above, although more complicated strategies may be pursued and even a more modest search space is very valuable.  Based on our target list and for an instrument with 1000 cm$^2$ and R = 3000 (nominal mission), this goal can be attained with $\approx $ 8 (1000 cm$^2$/A) (3000/R) Msec (Figure~\ref{fig:dNdz}).

One might consider a minimum mission with more modest instruments that only detect one-third of the absorption systems discussed above. It is sensible to use a higher threshold so that all of the time is not spent pressing down to low EW on just a few objects.  For an adopted a threshold of 5 m\AA\  (5$\sigma$), a collecting area of 300 cm$^2$ and a spectral resolution of R = 2000, we estimate that 33 absorption systems would be detected over a search space of $\sum z = 7.8$ and with a total exposure time of 8 Msec (Figure~\ref{fig:dNdz}).  This would constrain the normalization of $dN/dz$ to 17\%, but with poorer constraints on the location of the cutoff and the the shape of $dN/dz$.

Another goal would be to determine how $dN/dz$ changes as a function of redshift, which places important constraints on structure formation models for the past 8 Gyr.  This would require intentionally targeting objects at larger redshift, rather than observing objects ranked by our merit parameter.  There is a limit to the redshift distribution range because for $z \gtrsim 1$, the O VII line is shifted below 0.3 keV, where the combination of Galactic absorption and decreased sensitivity becomes quite significant, reducing the S/N of the spectrum.  It should be possible to determine at least the normalization of $dN/dz$ as a function of redshift to $z \approx 1$.  For an instrument with 1000 cm$^2$, R = 3000, this goal can be achieved but will approximately double the exposure times relative to the survey discussed above, but this depends on the shape of the instrumental sensitivity decrease below 0.4 keV. 

As simulations indicate a global increase in the hot gas fraction from z = 3 to the present day \cite{cen06}, there could be a significant decrease in the success of detecting absorption lines at redshifts approaching unity.  Therefore, a safer strategy would be to target the lower redshift AGN first ($z < 0.5$) before proceeding to higher redshift.

\subsection{The Distribution a Mass of Hot Halo Gas Around Galaxies}

Galaxies are likely to provide many of the best sites for X-ray absorption line studies.  The singular best case is the Milky Way, where O VII absorption has already been detected in about 20 sightlines \cite{miller13}.  The measured column densities along these sightlines, $3-10 \times 10^{15}$ cm$^{-2}$ (10-30 m\AA), are a factor of 5 times greater than for extragalactic sightlines because the density of gas is higher close to the Galaxy and our sightlines probe both the near and far halo.

To estimate the absorption properties of other galaxies similar to the Milky Way, we can project the Galactic halo onto the sky, and to do so, one needs the distribution of density and temperature as a function of radius.  We have determinations for these distributions based on the emission and absorption of the two oxygen ions.  Ref.\citenum{miller14} show that a gas density law of about $n \propto r^{-3/2}$ and T $\approx 2\times 10^6$ K fit both the O VII and O VIII emission line data, along with the O VII and O VIII absorption line data.  The data constrain the distribution to no further than 50 kpc, so at greater radii, the data require an extrapolation.  The observed absorption equivalent width in the anticenter direction is about 14 m\AA, ignoring optical depth corrections (making the somewhat uncertain corrections for optical depth will only increase the equivalent widths at large radii).  This provides a normalization to the absorption column density, and for simplicity, we adopt a power law in density to large radius ($n \propto r^{-3\beta}$).

This extended gas distribution does not include an exponential disk, which has been suggested for extended hot gas halos \cite{yao09b}.  With few sightlines, it can be difficult to discriminate between a disk and a spherical distribution \cite{fang13}, but with the many emission line sightlines \cite{henley12}, one has greater statistical power.  Ref. \citenum{miller14} found that an exponential disk did not fit the data but that a spherical distribution with a power-law decline in density was a successful fit; that is the model used here.  We have considered a combination of exponential disk and spherical halo model and the best fit indicates that the exponential disk is a 10\% contributor to the gas column (work in preparation), so we feel justified in neglecting it.

\begin{figure}
	\begin{center}
		\begin{tabular}{c}
			\includegraphics[height=7.cm]{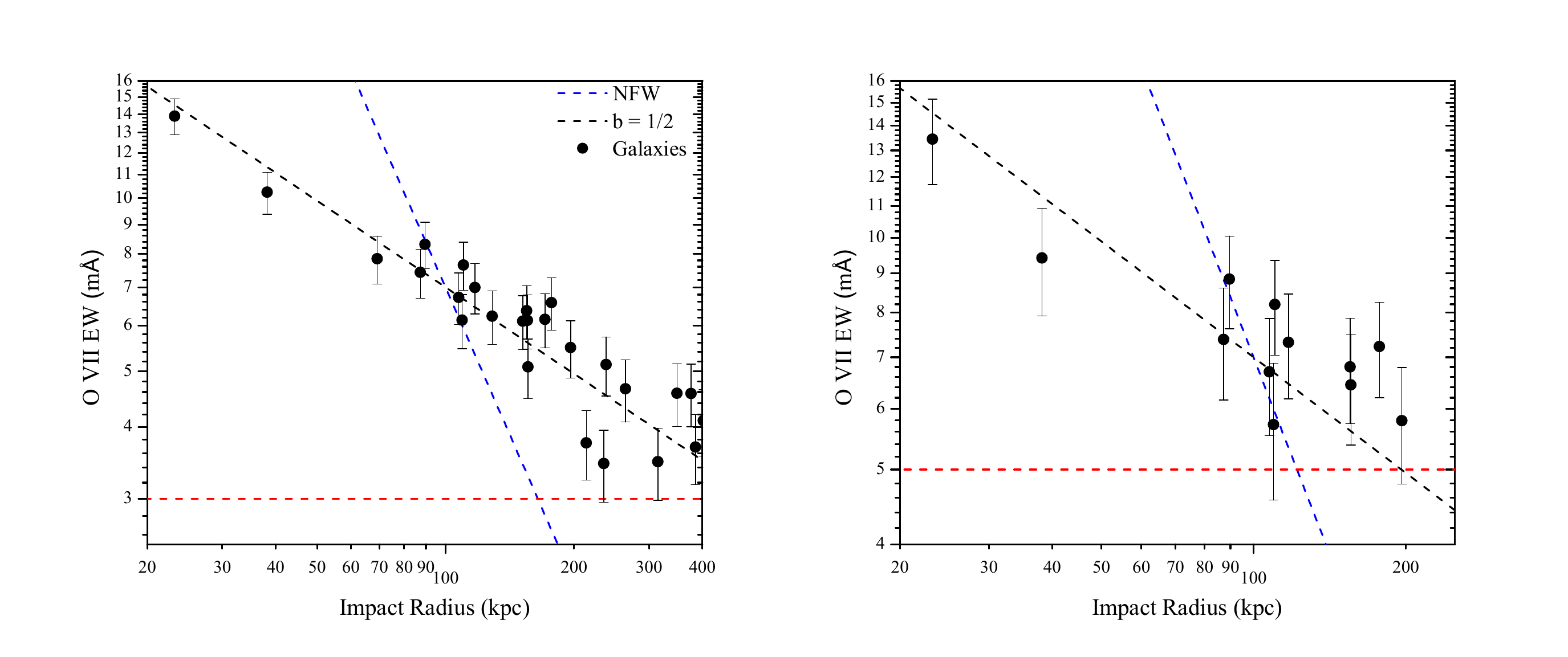}  
			\\
			(a) \hspace{5.1cm} (b)
		\end{tabular}
	\end{center}
	\caption{
		The absorption equivalent widths for sight lines through known galaxies (left panel), assuming that these optically luminous galaxies have the same absorbing halo as the Milky Way, for a density distribution $n \propto r^{-3/2}$ (black dashed line).  The blue dashed line is a NFW profile and the red dashed line that shows a 3 m\AA\ threshold in the EW.  The collecting area is 1000 cm$^2$ and with a resolution of R = 3000.
		The same sample (right panel), as observed with an instrument with a collecting area is 300 cm$^2$ and with a resolution of R = 2000.  The 5$\sigma$ threshold of 5 m\AA\ is shown as the horizontal red dashed line. 
	}
	\label{fig:EW_R_galaxies}
\end{figure}

Under the assumption that all galaxies brighter than $M_I \approx -20$ have a hot halo with the hot gas properties of the Milky Way, observing about 20 AGN targets of greatest merit plus selecting additional AGN near galaxies to a limiting EW of 3 m\AA\ ($\sigma$ = 0.5 m\AA ) will accurately define the density law to an uncertainty in $\beta$ of $0.03$ (Figure~\ref{fig:EW_R_galaxies}).  This could distinguish between a variety of models and a flattening at large radius would indicate the presence of an absorbing group medium.

For the less capable mission described above (effective area of 300 cm$^2$, R = 2000, with $\sigma$ = 1 m\AA ), about half of the absorption lines would be detected to an impact parameter of 200 kpc (Fig. 3b).  This would still permit one to distinguish between a NFW model, a $\beta$ = 1/2 model, and a flat model.

A measurement of the metal mass associated with the hot gas can be made, provided that one has a temperature estimate for the gas and there are a few ways of obtaining this.  One can estimate the temperature from $\beta$, which is the ratio of the virial temperature to the gas temperature, where the virial temperature can be inferred from the rotation velocity of a spiral galaxy.  Another approach is possible if the galaxy is close enough to detect in emission, in which case a temperature can be fit to the spectral energy distribution.  Such a temperature determination is weighted at radii less than 50 kpc, so one would have to extrapolate the temperature to larger radius. A third approach is to measure both the O VII and O VIII lines in absorption, where the ratio is sensitive to the temperature.  This requires an instrumental sensitivity sufficient to detect the O VIII line, which is about a factor of two weaker (for T = 2$\times$ 10$^6$ K), so that is likely detectable only for impact parameters less than about 150 kpc (for the telescope with A = 3000 cm$^2$).

The estimation of the total gaseous mass of the hot halo either requires adopting a value for the metallicity or relying on the Planck S-Z measurements.  The metallicity is calculated in simulations, with a typical mean mass-weighted value of $\sim 0.2$ of the solar value.  This is probably correct to a factor of two and gives a useful measure of the halo gas mass.  The other approach is where the S-Z signal from stacked galaxies is used to give a value for the gas mass responsible for the signal \cite{planckXI13,greco14}.  This heavily weights the hot gas halo but the typical galaxy, of luminosity L*, is below the level where the Planck signal is above the noise.  The Planck result would have to be extrapolated by a factor of five to get to L*, but given the tightness of the relationship, this may not cause great inaccuracies.  

There are various advantages to having greater sensitivity in that one can measure a variety of lines or even several lines in the same species.  For example, the ratio of the two O VII resonance lines at 21.60 \AA (He$\alpha$) and 18.63 \AA (He$\beta$) constrains the optical depth, which would be a modest concern if the lines are near the thermal value of the Doppler parameter (Figure 4).  As mentioned above, the ratio of the O VII to O VIII lines gives a measure of the temperature.  The O VII He$\beta$ line and the O VIII K$\alpha$ lines are about a factor of 3-5 times weaker than the O VII He$\alpha$ line, so for a limiting EW of 3 m\AA , the O VII He$\alpha$ line would have an equivalent width of 4-8 m\AA , so sightlines within about 60 kpc of a galaxy should lead to detections.  For an O VII EW of 13.9 m\AA\  and a Doppler b = 150 km s$^{-1}$, the optical depth at line center is 1, which leads to a modest column density correction. For a quiescent stationary halo, where the line width is the thermal Doppler width for oxygen (45 km s$^{-1}$), an optical depth of unity is reached for an EW = 4.2 m\AA\ .  The importance of optical depth effects and line shapes is the topic of a future paper.

\subsection{The Milky Way}

The study of hot gas absorption around external galaxies determines the column density as a function of radius by combining the results from different galaxies.  Multiple lines of sight through an individual galaxy provides much more detail and the two galaxies for which this can be accomplished are the Milky Way and M31.  Milky Way studies have proven fundamental to understanding every ISM component and a high S/N investigation of Galactic hot gas that will be of similar lasting value.  Such an investigation can reveal the shape and radial extent of the hot halo, its mass, metallicity, temperature, and the dynamics of this hot medium.  Because absorption is already measured, an order of magnitude sensitivity improvement will yield multiple galactic absorption lines from different levels and from several ions toward every target. 

\begin{figure}
	\begin{center}
		\begin{tabular}{c}
			\includegraphics[height=8.cm]{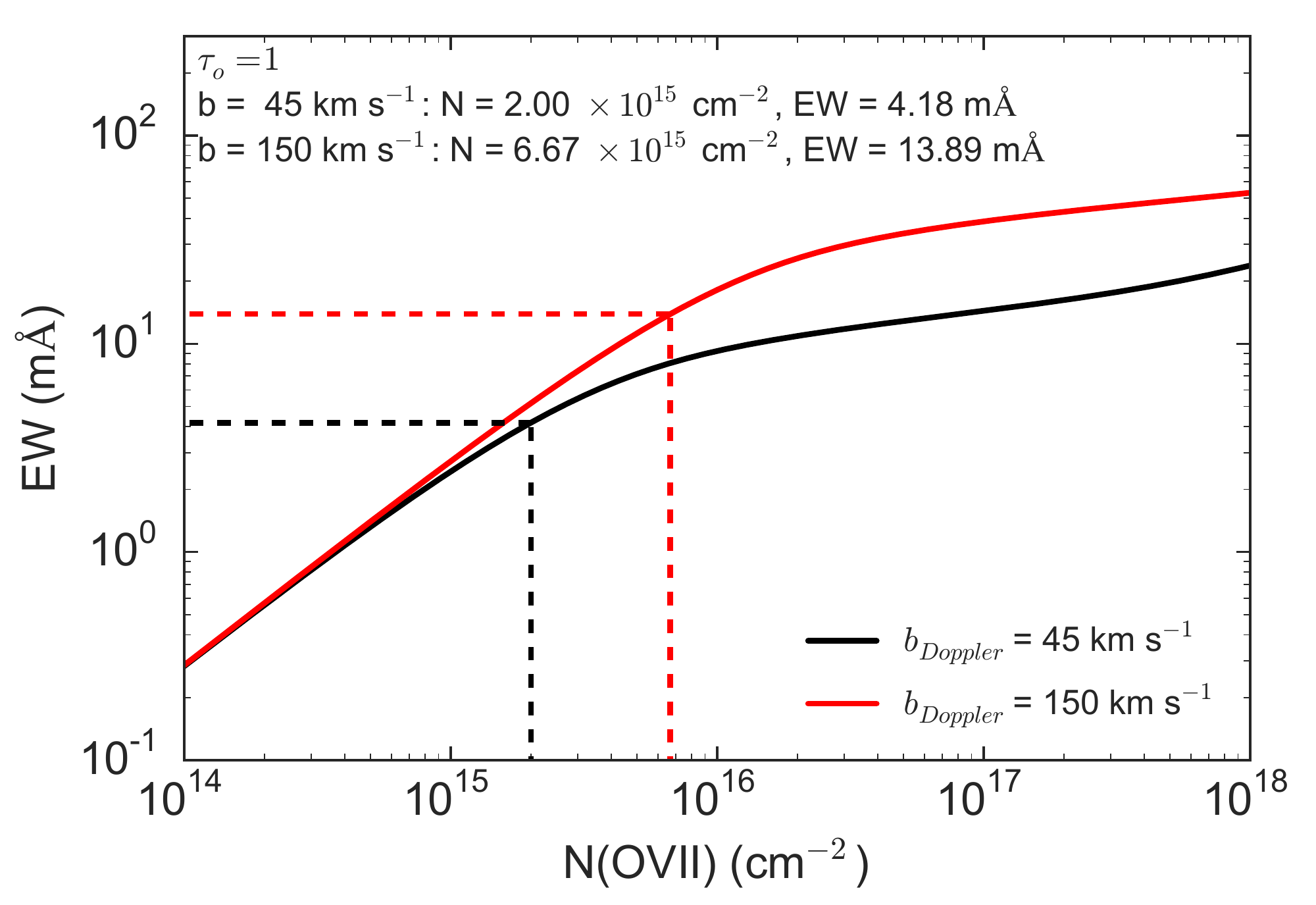}
		\end{tabular}
	\end{center}
	\caption{
		The curve of growth of the O VII He$\alpha$ line for a Doppler parameter at the thermal value for oxygen and for the thermal value for hydrogen, which would occur if there is modest turbulence. The red and black dashed lines show the columns and equivalent widths when the optical depth at line center is unity, which is when optical depth corrections need to be taken into account.
	}
	\label{fig:cog}
\end{figure}

Previous studies show that a hot halo exists around the Milky Way, with an estimate for the hot gas mass \cite{miller13,miller14}, and for three sight lines, a temperature estimate from the O VII to O VIII line ratios.  There are probably opacity considerations ($\tau \approx 1-4$ at line center; Figure~\ref{fig:cog}), but these are poorly constrained, yet they are potentially important as opacity corrections can raise the metallicity and change the inferred density and gas mass profile \cite{gupta12}.  Another current limitation is that most of the halo information is given by O VII absorption, which can be contaminated by the gas within and near the disk.  A potentially cleaner probe is the O VIII ion, for which opacity corrections are smaller, and contamination near the disk is relatively unimportant.

The observational goal is to measure the equivalent widths of O VII He$\alpha$, O VII He$\beta$, O VIII K$\alpha$, and O VIII K$\beta$, and to quantify the line shapes, given by the velocity centroid and line width (broadening parameter).  From the K$\alpha$ to K$\beta$ line ratio(s), we obtain the opacity, while the ratios of the adjacent species columns yield temperatures. A direct measure of the line width helps constrains the opacity, for lines of sight not substantially broadened by Galactic rotation, inflow, or outflow.  The average measured equivalent width of O VII He$\alpha$ is measured to be 20 m\AA\ \cite{miller13}, with the O VIII K$\alpha$ EW about 10 m\AA, so the anticipated values of O VII He$\beta$ is 4-6 m\AA.  For completeness, the O VIII K$\beta$ line would have a mean value of 3.5 m\AA, but this is a less important line as the opacity is more accurately measured using the two O VII lines.  The three most important lines are expected to have mean values in the 7-20 m\AA\ range, which is significantly less demanding than either the galaxy halos requirements or the external galaxy halos requirement. Consequently, observations made for those programs can double in use for a Milky Way program.

\begin{figure}
	\begin{center}
		\begin{tabular}{c}
			\includegraphics[height=12.cm]{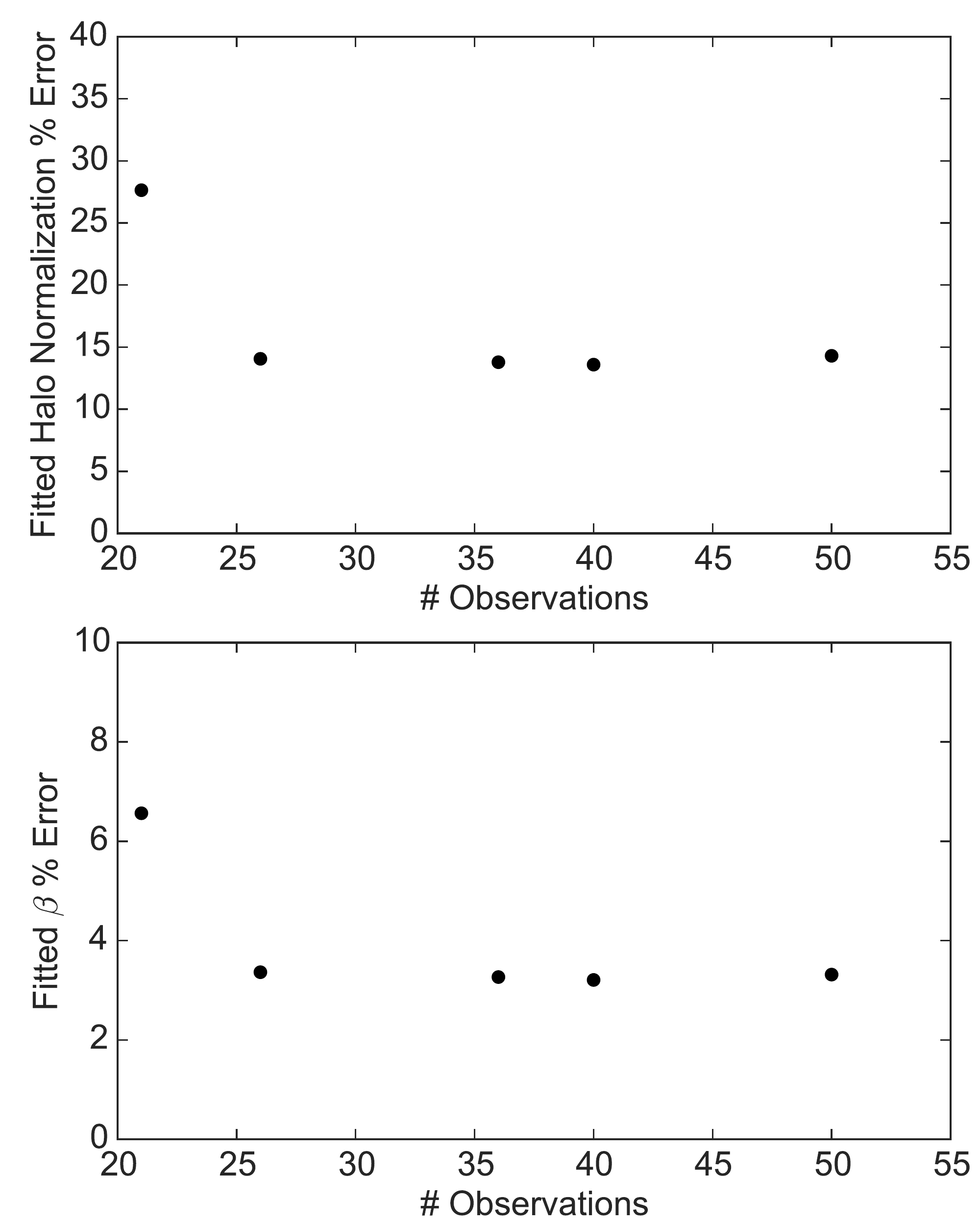}
		\end{tabular}
	\end{center}
	\caption{
		The uncertainties in fitting for a power-law density distribution, $n = A r^{-3\beta}$, as a function of the number of lines of sight.  We added a random variation between lines of sight of 4 m\AA , where the median EW is 16 m\AA .
		After the first 21 simulated observations, randomly placed on the sky, five targets were added going across the Galaxy, reducing the uncertainty significantly.  Adding further randomly located lines of sight leads to only modest improvements.
	}
	\label{fig:MW_beta_A}
\end{figure}

We carried out simulations to determine how the number of sight lines affects the constraints on the density distribution.  
We adopted a parametric model for the density with the form $n(r) = n_o (1+(r/r_c)^2)^{-3{\beta}/2}$ and found that it is extremely important to have a number of sight lines go across the Milky Way. 
Practically, that means there should be at least five sightlines within about 35$^{\circ}$ of the Galactic center, and as this constitutes only one-tenth of the sky, there is a significant advantage to add such sight lines to modest size samples.  
To illustrate this point, we simulated fits where the EW along each sightline has an uncertainty of 10\%.  For a random sample on the sky with 21 sightlines of high merit, the uncertainty in the density normalization is 28\% and for $\beta$ it is 6.5\% (Figure~\ref{fig:MW_beta_A}).  When we add five sightlines across the Galaxy, within 35$^{\circ}$ of the Galactic center (from within the top 70 targets), the uncertainties drop to 14\% and 3.5\%.  Further addition of random sightlines hardly improves the uncertainties.

For this program, one can use closer AGNs than for intergalactic studies, as there is no need to search redshift space to measure Milky Way properties.  
Therefore, if only a Milky Way study were being proposed, the targets would be rank-ordered by flux, with little consideration of redshift.

Since existing instrumentation already has made important contributions, the requirements for this project are more modest than the galaxy halos or large scale structure programs.  The most important improvement should be in spectral resolution, which will help determine the dynamics of the absorbing material and help to separate Galactic from Local Group components.  For an observing time of 10 Msec and a spectral resolution of 3000, the collecting area only needs to be $\sim 100$ cm$^2$ to obtain 35 targets.

\begin{figure}
	\begin{center}
		\begin{tabular}{c}
			\includegraphics[height=12.cm]{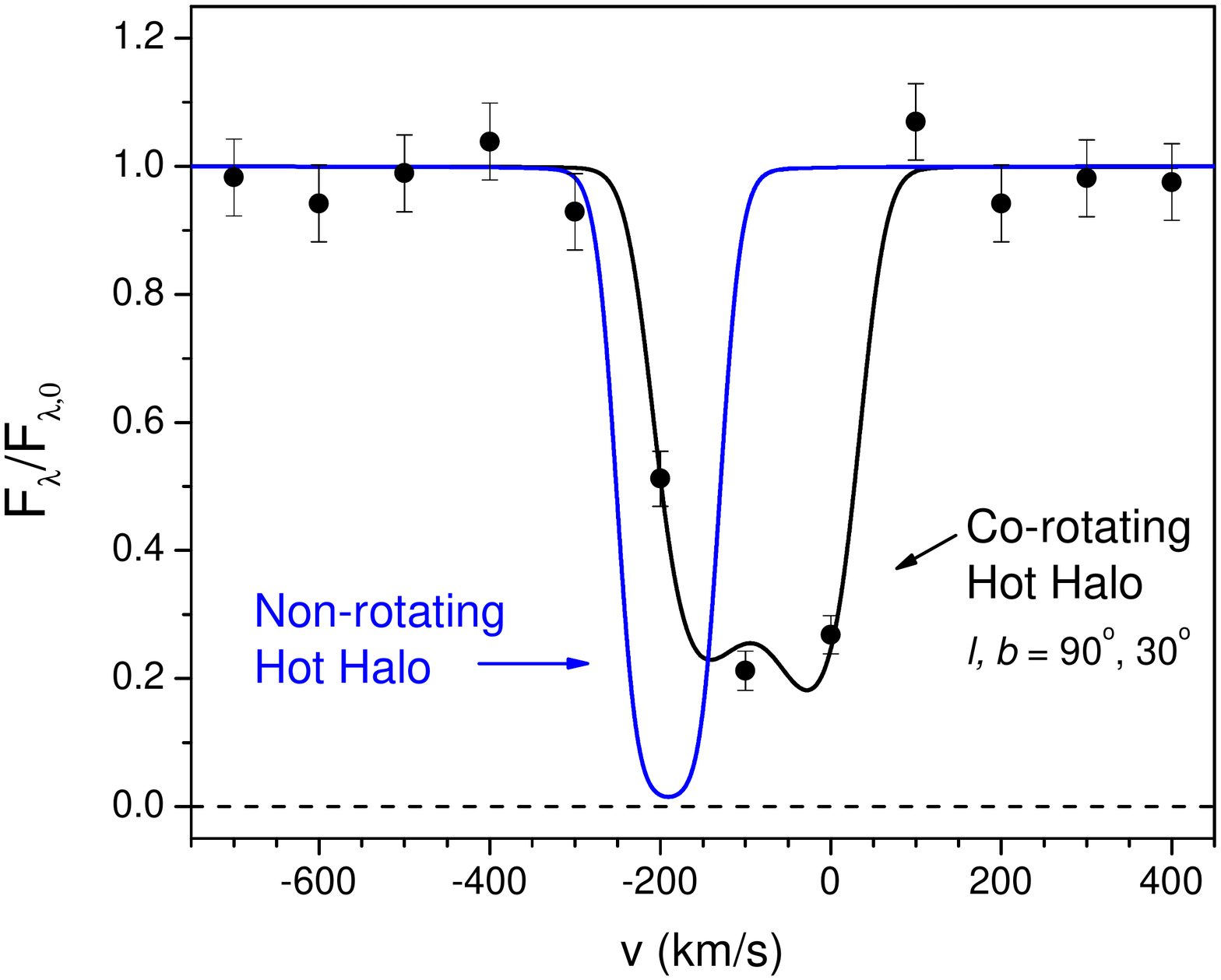}
		\end{tabular}
	\end{center}
	\caption{
		The O VII He$\alpha$ absorption against a normalized continuum, showing two models: a stationary halo (blue) and a co-rotating halo (black) in the same direction, $l, b = 90^{\circ} , 30^{\circ} $ .  A simulated spectrum for the co-rotating halo, shown as black points for a resolution of 100 km s$^{-1}$ (R = 3000) and a total EW for the line determined to 0.5 m\AA .  The two models are easily distinguished and the optical depth corrections are both significant and different.
	}
	\label{fig:MWrot}
\end{figure}

The above simulations were for fits to distributions of the equivalent widths under the assumption that optical depth effects are not important.  However optical depth effects are important, as are the rotational properties of the Galaxy and hot halo.  This is reminiscent of 21 cm studies of neutral hydrogen in the Milky Way, and the velocity equations are very similar.  This topic will be discussed in detail in a separate work, but we demonstrate some of the issues by considering a single sight line at $l, b = 90^{\circ} , 30^{\circ} $.  If the halo is stationary relative to the rotating disk, the line is shifted by $-v_{rot} ~cos(b)$ and it has a near-Gaussian shape.  For the density inferred from Ref.\citenum{miller14}, and for a quiescent halo where the Doppler b value has the thermal width for oxygen (45 km s$^{-1}$), the EW = 8 m\AA\  and the optical depth at the line center is 4 (Figure~\ref{fig:MWrot}).  If the halo gas is co-rotating with the disk, for the same value of the Doppler b parameter, the line is broadened by about a factor of two due to rotational effects.  This lowers the optical depth across the line, producing a line with an EW = 14 m\AA\, about twice the value of the non-corotating case (for a constant column), but still lower than the EW if absorption effects had been unimportant (24 m\AA ).  Clearly, optical depth corrections will need to be made, and to do this in a unique fashion, one needs good spectral resolution and high S/N.
For the case illustrated, a velocity resolution of 50 km s$^{-1}$ would be helpful, corresponding to a resolution of R = 6000.  However, if there is significant turbulent broadening (comparable to the sound speed of the gas), a resolution of 2000-3000 would be adequate.

\subsection{Messier 31}

The other galaxy of major importance is Messier 31, which is the only massive external galaxy with multiple lines of sight through the halo.  With $v_{Helio}$ = -300 km s$^{-1}$, it is separated in velocity from the Milky Way.  There are six X-ray bright AGNs between 100-200 kpc, one of which would naturally be observed in the large scale structure program.  The other five targets are a few times fainter, so for an instrument with R = 3000 and 1000 cm$^2$, it would require about 1 Msec to detect absorption from O VII He$\alpha$ with values of about 3 m\AA.  Such a study would give information about the mass of hot enriched gas in the halo and whether there is a metallicity gradient within the halo.

We summarize the science outcomes and observing time required for these various studies in Table 5.

The LMC and SMC are not suitable for X-ray absorption line studies.  They lie well within the Milky Way halo, so isolating the absorption strictly associated with those galaxies is not possible.  Also, the mass of these galaxies is well below the value for which $10^6$ K gaseous halos are predicted to naturally develop.

\subsection{Improving Observing Efficiency Through Variability}

Employing the natural variability of AGNs can improve the effectiveness of the observing program either by obtaining a higher S/N for fixed observing time, or observing additional objects.  The typical variability of a target in our sample is about 30\% (1$\sigma$), although outbursts can lead to significantly higher flux densities.  For a sample of six targets, one would lie at least 1$\sigma$ above the mean two-thirds of the time and at least one would lie above their mean 98.5\% of the time.  The ability to identify a particularly bright source improves quickly with the number of objects, yet with observing angle restrictions and the different merit values of the objects, it would be difficult to consider a large number of targets.  By monitoring the most desirable targets prior to observation, one constructs an instantaneous value of the merit of different candidates.

The monitoring is most accurately obtained through X-ray observations, provided that a telescope exists that is capable of making such observations.  The X-ray continuum can vary significantly and on timescales as short as hours for some blazars (e.g., Ref.\citenum{mchardy14, isobe15}).  However, aside from outbursts, the typical periods of brightening and dimming are generally significantly longer, being weeks to months.  This can be compared to the typical length of a spectroscopic observation, expected to be 1-20 days of clock time (0.1-1 Msec of observing time).  Since the observing times are often shorter than the time for AGNs to change their X-ray flux, this monitoring strategy can be effective.

While X-ray monitoring is preferred, there is a strong correlation between X-ray variability and that at other wavelengths.  In particular, there is a very good correlation between X-ray and optical brightness variations (e.g., Ref.\citenum{cameron12,mchardy14}), with lags that are short (hours) compared to the length of the X-ray exposure times (days).  The targets could be monitored with only modest size telescopes, and some of the optically brighter AGNs are regularly monitored by amateur astronomers.  In addition to modest size optical telescopes at traditional public and private observatories, there are some world-wide telescope networks that can perform these tasks (e.g., the Las Cumbres Observatory). 

\subsection{Instrumental Requirements}

The observational needs of executing a science plan translates directly into instrumental requirements.  One consideration is the required S/N, which is closely related to the resolution of the instrument.  In calculating this, we use the strongest line, expected to be the 21.60 \AA\ O VII resonance line in a gas at $2 \times 10^6$ K.  If there is no turbulence or bulk motions, the intrinsic FWHM of this line would be 45 km s$^{-1}$, while if there is bulk motion or turbulence of order of the sound speed of the gas, the FWHM is expected to have a value of about 200-300 km s$^{-1}$.  To resolve the lines a resolution of at least 1,500 is required and observations will benefit by higher resolution, possibly as high as a resolution of 10,000, which would oversample lines.

For an unresolved O VII line with an equivalent width of 3 m\AA\ and at low redshift, the characteristic depth of the line, as a fraction of the continuum is 40\% for a resolution of 3000.  If this is a 5$\sigma$ detection, then the requirement on the S/N due to the intrinsic properties of the instrument is about S/N $>$ 25, which is easily accomplished with modern detectors.  At significantly lower resolution, the challenges to detector stability are much greater.  For the case of a quantum microcalorimeter detector, such as the X-IFU on \textit{Athena}, the resolution for low redshift O VII lines is about 300 (assuming 2 eV resolution), a figure that becomes worse when the lines are redshifted to lower energies.  For a resolution of 300, the characteristic depth of the line is 4.2\% and a 1$\sigma$ value of 0.8\% for a 5$\sigma$ detection.  The instrument must be stable enough to obtain a S/N of about 200 in order not to significantly degrade the line measurement.  This requirement places significant demands on the construction of the instrument.

There are other advantages to higher spectral resolution in terms of scientific gains.  The UV absorption line systems are often highly structured and we might expect structure in the X-ray regions as well.  One example of this was shown for a co-rotating hot Galactic halo (Figure~\ref{fig:MWrot}).  Another Galactic phenomenon is a fountain, where the gas goes up hot and comes down cool, in which case one would see a redshift of the lines due to the hot rising gas, most prominent toward high latitude sources.
For external galaxies, the presence of a large-scale galactic wind can lead to shells of hot gas that occur as gas piles up near a contact discontinuity with a confining medium, as is found in cosmological simulations.  This will lead to a double-line with absorption redshifted and blueshifted relative to the systemic velocity of the galaxy.  The expected velocities might be estimated from the characteristic potential well depth of galaxies and groups, being 200-1000 km s$^{-1}$.  A resolution better than 200 km s$^{-1}$ is needed to study these systems, which implies a R $>$ 1500.  Also, improved resolution permits one to more accurately identify the velocity of the line center, which is valuable when comparing to cooler gas and to the stellar content of nearby systems.

\section{Discussion and Conclusion}

There have been a number of IGM detections published and several upper limits presented, so we examine whether they are consistent with the expectations for individual galaxies and from structure function predictions.

\subsection{Should Extragalactic Absorption Have Been Detected Already?}

Based on the predictions of $dN/dz$ \cite{cenfang06}, we can compare the number of absorption systems expected for the few sources with sufficiently small equivalent width uncertainties.  For Mrk 421, the weighted uncertainty, combining the \textit{Chandra} and \textit{XMM-Newton} data is 0.5 m\AA, so a 4$\sigma$ limit would be 2 m\AA, for which $dN/dz$ = 10.  However, the redshift is only 0.03, so 0.3 absorption systems would be predicted, a value that is reduced if we exclude the velocity region around the redshift of Mrk 421, as AGNs can produce their own local features due to winds.  Two other bright objects with larger redshifts are PKS 2155-304 (z = 0.116), where $\sigma$(EW) = 1.2 m\AA\ for the XMM data set and 3C 273 (z = 0.158), where where $\sigma$(EW) = 2 m\AA\ for the XMM data.  At a 4$\sigma$ threshold, one would expect 0.6 absorption systems for PKS 2155-304 and 0.4 absorption systems in 3C273, but none are reported.  These are the best targets as other possible AGNs are either fainter, have less observing time, or have low redshifts.  Taken together, one would expect 1.2 absorption systems with EW in the range 2-6 m\AA, not excluding any redshift regions around these objects. 

X-ray absorption features have been found at the 4-7$\sigma$ levels, with the two best detections being the O VII He$\alpha$ resonance line at z = 0.03 (22.3 \AA) in the X-ray continuum of both Mrk 421 and H2356-309, the latter being through the Sculptor Wall \cite{nica02,fang10}.  However, Ref.\citenum{nica14} has recently pointed out that this feature at 22.3 \AA\ is not only seen in other AGN, it is seen in the nearby X-ray binary LMC X-3.  He argues that this must be a Galactic absorption feature and suggests an identification with an inner shell oxygen absorption line and not a redshifted intergalactic O VII line.

Each of these three AGNs have a galaxy closer than 300 kpc projected on to the sky, but in each case, the galaxy is about at the same redshift as the AGN, so it is possible that the galaxy is behind the AGN rather than in front of it.  The galaxy SDSS J110428.26+381240.1 has a projected impact parameter of only 9.1 kpc from Mrk 421 but there is no evidence for absorption in the sensitive UV lines (Danforth et al. 2011), so either it is  behind the galaxy or the ionizing radiation from the BL Lac object has strongly ionized the gas so that there are no viable absorbers.

Similarly, for 3C 273, there is a galaxy at about the same redshift and at a projected distance of 238 kpc (SDSS J122911.67+020312.7), yet there are no UV absorption lines.  We would predict an O VII absorption line strength of 4-5 m\AA, which is below even a 3$\sigma$ threshold with existing data.  Even if this intervening galaxy were in front of 3C 273, it would not produce a statistically significant O VII absorption line.

The other AGN, PKS 2155-304, lies in a group of galaxies, with a moderately bright galaxy (M$_R$ = -20.29) at a projected distance of 110 kpc.  If it were a foreground galaxy that is not highly ionized by the AGN, it would produce a O VII absorption line of $\approx$7 m\AA, producing about a 5$\sigma$ detection with existing data.  However, no such line is seen nor are there significantly strong UV absorption lines at this redshift, so either it is not a foreground galaxy, or its gas is highly ionized.  Ref.\citenum{fang02} claim a 4.5$\sigma$ O VIII K$\alpha$ absorption line using the Chandra LETG, attributed to an intervening galaxy group at lower redshift, yet higher S/N data from XMM fail to confirm the detection \cite{rasm03}. 

One possible site for X-ray absorption lines is from the galaxy halo of the AGN host system.  This approach has the natural problem that it is impossible to distinguish between an absorption feature associated with the environment close to the AGN and a galactic halo.  Several low luminosity AGN have many strong X-ray features (including O VII; e.g., Ref.\citenum{kaastra11}), which are usually attributed to winds or accretion (e.g., Mrk 509, MCG -6-30-15).  However, many AGNs fail to show O VII absorption features \cite{yao09} even though the line would be easily detected if the AGN was surrounded by a Milky Way-like hot halo.  The failure to detect such lines is likely attributable to the powerful ionizing radiation and heating of the local environment by a luminous AGN.

The difficulty with detecting X-ray absorption lines is a combination of limited search space and poor S/N, so by stacking spectra, one can improve the S/N of absorption features.  This approach was taken by Ref.\citenum{yao10}, who aligned X-ray spectra with the redshifts of known galaxies along multiple lines of sight but failed to detect absorption features.  Their upper limit of about 2 m\AA\ (5$\sigma$) per 20 m\AA\ search bin would seem to be quite restrictive, so we reexamined their approach.  Included in their stacking study are mostly low luminosity galaxies, which theory tells us are unlikely to produce hot halos that could create O VII absorption.  Also, many of these low luminosity galaxies are projected at distances greater than their virial radius, which mitigates against hot gas production.  If one restricts the list to optically luminous galaxies that might produce hot halos (M$_r < -20$), for projected separations less than their virial radii, and for galaxies clearly below the redshift of the AGN, there are few galaxies to stack and the failure to detect an O VII absorption line is consistent with predictions.

\subsection{The Future of Existing Observatories for Absorption Line Studies}

The envisioned instrumental configurations, even if approved in the near future, would not be operational until the 2020s, so we consider whether any aspects of the absorption line studies might be carried out with existing instruments.
The study of the Milky Way hot gas was established with \textit{XMM-Newton} (primarily) and \textit{Chandra}.  Since the RGS is always operating on \textit{XMM-Newton}, it has obtained spectra for every object it observed.  For the brightest sources, a great deal of observing time was accumulated, leading to about two dozen sources with useful Galactic O VII absorption lines.  There are no very bright sources without substantial observing time, so one would have to observe less bright sources.  With fainter sources, to significantly improve the sample size would require $>$ 30 Msec, which is impractical.

For the extragalactic absorption goals, one strategy is to choose the best target over the effective redshift range of the RGS, which for O VII is 0 $\leq$ z $<$ 0.73.  The collecting area is about 43 cm$^2$ at 21.6 \AA\, but above 24 \AA\ (z = 0.11), it increases to about 90 cm$^2$, steadily decreasing to about 40 cm$^2$ at 37 \AA\ (z = 0.71), after which it declines quickly.  The resolution increases steadily as R = $360 ~(1+z)$, which aids in line detection as the redshift increases.
The highest merit target that covers a significant redshift space is 1RXS J153501.1+532042, at z = 0.8900, and F (0.5-2 keV) = 9.57$\times$10$^{-12}$ erg cm$^{-2}$ s$^{-1}$.  We choose an observational target of detecting 4 absorption line systems, which only gives a rough notion of the normalization of $dN/dz$, as the 90\% limits on the Poisson distribution are 1 - 6 absorption systems.  However, there is only a 2\% chance of obtaining zero detections (for a mean of 4), so a positive result should be obtained.  When we require that detections be at the 5$\sigma$ level or above, a total integration time of 15 Msec is required, provided that there are no systematics in the detector (only statistical uncertainties are used).  If one were willing to accept a 4$\sigma$ detection threshold, the exposure time is reduced to 10 Msec.  A project of this nature cannot be carried out with the RGS with significantly less time.  Also, great care would have to be taken to understand the systematics, as the line depth would be about 7\% of the continuum, so a 1$\sigma$ statistical value would be 1.5\%, which means that the systematic errors should be less than 1\% in order not to degrade a 5$\sigma$ detection to below a 4$\sigma$ detection.  Finally, if one were to use the LETG on \textit{Chandra}, the exposure times would be several times longer, primarily due to the lower effective collecting area (Table 1).

Another and perhaps better strategy for studying intergalactic absorption lines, inferred from the shape of $dN/dz$, is to observe many objects to shallow depth and to maximize $\sum z$ with a limiting EW near the turnover in $dN/dz$.  The goal would be to obtain the normalization of $dN/dz$ at high EW.
A good value for the limiting EW is 10 m\AA\ (at 21.6 \AA ), and if one is to detect this at the 4$\sigma$ level, one would require $\sigma$ = 2.5 m\AA\ with the RGS.  For EW $\gtrsim$ 10 m\AA\ , there is about one O VII absorption system per unit redshift, so the best objects are those at redshifts 0.3-0.7 (the upper limit set by the RGS limit), and the exposure times for the best objects are typically 1-2 Msec. To accumulate five detections, one would observe the best nine targets, reaching $\sum z \approx$ 4.1, which requires about 10 Msec.  If the Ref.\citenum{cen12} $dN/dz$ distribution is correct, one would detect about 10 absorption systems.  Doing this with 5$\sigma$ detections per absorption line would entail about 15 Msec.  This would constitute an extremely large program, but it could be carried out over a few years if deemed sufficiently important.

\subsection{The Relative Importance of Galaxies, Groups, and Clusters}

\begin{figure}
	\begin{center}
		\begin{tabular}{c}
			\includegraphics[height=10.cm]{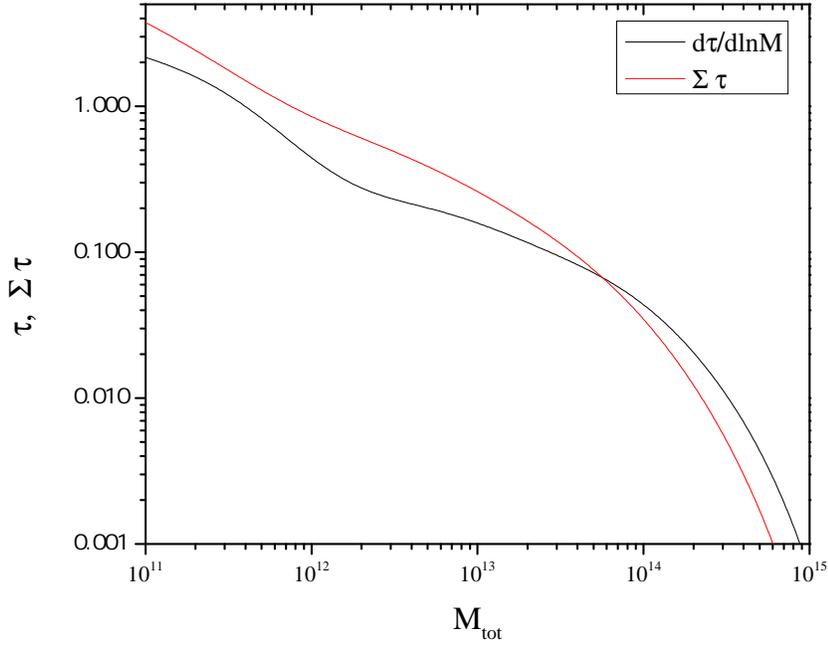}
		\end{tabular}
	\end{center}
	\caption{
		The differential optical depth as a function of mass (black line) and the cumulative optical depth above some mass (red line), for a path length of z = 0.3.  There is no correction for the production rate of hot gas as a function of mass or for ionization corrections.  Galaxy groups and clusters dominate above a mass of $\sim 10^{12.5}-10^{13}$  M$_{\odot}$ and galaxies dominate at lower masses.  This shows that the lower mass groups will be more important in producing absorption features than galaxy clusters.  Individual galaxies would appear to be more important than galaxy groups, but theory suggest that galaxies below $\sim 10^{12}$ M$_{\odot}$ may not be efficient at producing hot gaseous halos.
	}
	
	\label{fig:dtau_dM}
\end{figure}

The likely sites for X-ray line absorption are provided in part through simulations of large-scale structure and the observed properties of hot gas around the Milky Way.  Further insights are possible when we consider the space densities of virialized systems as a function of mass.  This function has been calculated by a variety of authors and here we adopt the work of Ref.\citenum{tinker08}.  For each mass, we calculate a virial radius so that the cross section of detectability is just $\pi \eta R_{vir}^2$, where we adjust $\eta$ such that the result agrees with the simulations of Ref.\citenum{cenfang06}; we find $\eta \sim 2$.  As shown in Figure~\ref{fig:dtau_dM}, absorption by high mass clusters will be uncommon, mainly because of their relatively low space density.  For lower mass clusters and galaxy groups ($10^{13}-10^{14}$ M$_{\odot}$), the optical depth due to absorption rises slowly toward lower mass systems.  It then rises quickly toward galaxy masses, but missing from this simple calculation is the rate of production of hot gas as a function of system mass.  Nevertheless, it points out the need to know about galaxy groups and galaxies along lines of sight.  Published galaxy group catalogs generally do not yet include systems in mass ranges below $10^{14}$ M$_{\odot}$, but future optical surveys should be able to fill this in and provide us with excellent information about the best sites for X-ray absorption lines.

We have only limited knowledge of the mass, metallicity, spatial structure, and dynamics of the hot component of the universe, yet we can take great strides with modest instrumentation dedicated to this task.  Current instrumentation on \textit{Chandra} and \textit{XMM} has a resolution of about 800 km s$^{-1}$ at the O VII and O VIII lines, about an order of magnitude larger than the Doppler width of the lines.  These same instruments have collecting areas between a large postage stamp and a credit card, so there is ample room for improvement.  \textit{Astro-H} will have more collecting area, but such poor spectral resolution (R = 100; 3000 km s$^{-1}$) that it is unable to detect any lines except strong Milky Way lines (a typical IGM line would have a line depth of only 1.5\%, so a S/N exceeding 500 would be needed). 

Within the \textit{NASA} Explorer or Discovery class missions, for example, it should be possible to improve on the product of spectral resolution and collecting area by one to two orders of magnitude.  This should lead to detections of hot absorbing gas from galaxies and galaxy groups that will not only complete the census of gas and metals in the Universe, it will reveal how hot mode accretion and hot gas production occurs in galaxies.  A failure to commonly detect this gas would be in conflict with the S-Z results, the hot gas absorption and emission from the Milky Way, and the failure to find more than 10\% of the metals in cooler gaseous phases.  This is the last significant gaseous phase of the Universe waiting to be explored and we possess the technology and resources to make major discoveries.

\subsection{Acknowledgments}
We would like to thank Jon Miller, Randall Smith, Fabrizio Nicastro, Eric Miller, Renyue Cen, Chris Miller, Jelle Kaastra, Mike Anderson, and Frits Paerels for the insights and suggestions that they provided.  The authors gratefully acknowledge NASA Award NNX11AJ55G for financial support of this work.
This research has made use of data, software and/or web tools obtained from the High Energy Astrophysics Science Archive Research Center (HEASARC), a service of the Astrophysics Science Division at NASA/GSFC and of the Smithsonian Astrophysical Observatory's High Energy Astrophysics Division.
Some of the optical images and redshifts for this work were made possible by SDSS-III, where funding has been provided by the Alfred P. Sloan Foundation, the Participating Institutions, the National Science Foundation, and the U.S. Department of Energy Office of Science.
Some of the optical images discussed in this paper were obtained from the Mikulski Archive for Space Telescopes (MAST). STScI is operated by the Association of Universities for Research in Astronomy, Inc., under NASA contract NAS5-26555. Support for MAST for non-HST data is provided by the NASA Office of Space Science via grant NNX13AC07G and by other grants and contracts.


\bibliography{report}   
\bibliographystyle{spiejour}   

\newpage

\begin{deluxetable}{lllrrrrrr}
\tabletypesize{\scriptsize}
\tablecaption{Operating, Planned, and Notional Instruments}
\label{tab:instruments}
\tablewidth{0pt}
\tablehead{\colhead{Observatory} & \colhead{Instrument} & \colhead{Type} & \colhead{E$_{low}$} & \colhead{E$_{high}$} & \colhead{R} & \colhead{A$_{eff}$}  & \colhead{R~$\times$~A$_{eff}$} & \colhead{EW Limit} \\
& & & \colhead{keV} & \colhead{keV} & \colhead{0.5 keV} & \colhead{cm$^2$} & \colhead{10$^3$} 
}
\startdata
Chandra	&	LETG/ACIS-S	&	grating	&	0.21	&	10	&	500	&	3.3	&	1.65	&	86.7  \\
XMM	&	RGS	&	grating	&	0.33	&	2.5	&	420	&	45	&	18.9	&	25.6 \\
Nominal	&	this paper	&		&	0.2	&	1.5	&	3000	&	1000	&	3000	&	2.0 \\
Minimum	&	this paper	&		&	0.2	&	1.5	&	2000	&	300	&	600	&	4.5 \\
Athena	&	X-IFU	&	calorimeter	&	0.2	&	12	&	250	&	8200	&	2050	&	2.5  \\
\enddata
\tablecomments{The resolution and effective area for existing and planned missions are those listed in mid-2015. The EW limit is for a 4$\sigma$ detection, a 0.3 Msec exposure, and a 0.5-2.5 keV flux of 1$\times$10$^{-11}$ erg cm$^{-2}$ s$^{-1}$, corresponding to an integrated flux of 3$\times$10$^{-6}$ erg cm$^{-2}$. The effective area of the XMM RGS changes rapidly near 0.5 keV (the RGS2 contributes), so we used the lower value at slightly higher energies.  A resolution of 2 eV is used for the Athena X-IFU.
}
\end{deluxetable}

\begin{deluxetable}{lrrrrrrrccr}
\tabletypesize{\scriptsize}
\tablecaption{Ions and Absorption Lines}
\label{tab:ions}
\tablewidth{0pt}
\tablehead{\colhead{Ion} & \colhead{$\lambda$} & \colhead{E} & \colhead{f$_{ik}$} & \colhead{[X/H]} & \colhead{Rel O VII} & \colhead{log$T_{min}$}  & \colhead{log$T_{max}$} & \colhead{MW} & \colhead{IGM} & \colhead{$z_{max}$} \\
& \colhead{\AA} & \colhead{keV} & 
}
\startdata
C V He$\alpha$	 & 	40.26	 & 	0.308	 & 	0.647	 & 	8.59	 & 	1.24	 & 	5	 & 	6.1	 & 	X	 & 	X	 & 	0.23	  \\ 
C VI Ly$\alpha$	 & 	33.73	 & 	0.368	 & 	0.416	 & 	8.59	 & 	0.39	 & 	5.75	 & 	6.35	 & 	X	 & 	X	 & 	0.47	  \\ 
N VI He$\alpha$	 & 	28.79	 & 	0.431	 & 	0.674	 & 	7.93	 & 	0.20	 & 	5.2	 & 	6.3	 & 	X	 & 		 & 	0.72	  \\ 
N VII Ly$\alpha$	 & 	24.80	 & 	0.500	 & 	0.416	 & 	7.93	 & 	0.057	 & 	5.9	 & 	6.6	 & 	X	 & 		 & 	1.00	  \\ 
O VII He$\alpha$	 & 	21.60	 & 	0.574	 & 	0.695	 & 	8.74	 & 	1.00	 & 	5.4	 & 	6.5	 & 	X	 & 	X	 & 	1.30	  \\ 
OVIII Ly$\alpha$	 & 	18.97	 & 	0.654	 & 	0.416	 & 	8.74	 & 	0.25	 & 	6.1	 & 	6.8	 & 	X	 & 	X	 & 	1.62	  \\ 
Fe XVII	 & 	15.02	 & 	0.826	 & 	2.310	 & 	7.45	 & 	0.08	 & 	6.25	 & 	7	 & 	X	 & 		 & 	2.30	  \\ 
Ne IX He$\alpha$	 & 	13.45	 & 	0.922	 & 	0.721	 & 	8.00	 & 	0.12	 & 	5.7	 & 	6.9	 & 	X	 & 		 & 	2.69	  \\ 
Fe XX	 & 	12.82	 & 	0.967	 & 	0.520	 & 	7.45	 & 	0.01	 & 	6.75	 & 	7.2	 & 		 & 		 & 	2.87	  \\ 
Mg XI He$\alpha$	 & 	9.169	 & 	1.352	 & 	0.741	 & 	7.54	 & 	0.03	 & 	5.95	 & 	7.1	 & 		 & 		 & 	4.41	  \\ 
Si XIII He$\alpha$	 & 	6.648	 & 	1.860	 & 	0.747	 & 	7.54	 & 	0.02	 & 	6.2	 & 	7.25	 & 		 & 		 & 	6.44	  \\ 
\enddata
\tablecomments{The oscillator strength, f$_{ik}$, is given in the fourth column. The sixth column is a guide to the relative absorption line strength and is the product of the oscillator strength, wavelength, abundance, and the maximum ion fraction possible for the ion. $T_{min}$ and $T_{max}$ are the temperatures where the ion falls to one-tenth its maximum value, indicating the useful temperature range of the ion. The most likely lines used for Milky Way (MW) and extragalactic studies (IGM) are denoted in columns 9, 10, while the maximum practical redshift is given in the last column.
}
\end{deluxetable}

\begin{deluxetable}{cllllrccccrr}
\tabletypesize{\scriptsize}
\rotate
\tablecaption{\label{tbl-1}}
\tablecaption{Background AGN Targets}
\label{tab:objects}
\tablewidth{0pt}
\tablehead{
\colhead{\#} & \colhead{Name} & \colhead{Alt name} & \colhead{RA} & \colhead{DEC} &
\colhead{redshift} & \colhead{SDSS} & \colhead{HST} &
\colhead{Galaxy} & \colhead{Group} &
\colhead{Flux} & \colhead{Merit}
}
\startdata
1   & PKS 2155-304             &                               & 329.7169379                         & -30.2255883             & 0.116                        & N    & N   & 2      &       & 1.85E-10                      & 2.14E-11                     \\
2   & Ton 116                  & 1RXS J124312.5+362743         & 190.8030833                         & 36.4622222              & 1.0654                       & Y    & N   & 1.5      &       & 8.97E-12                      & 8.97E-12                     \\
3   & 1RXS J153501.1+532042    & 1ES 1533+535                  & 233.75334                           & 53.343703               & 0.89                         & Y    & Y   & 1      &       & 9.57E-12                      & 8.52E-12                     \\
4   & PG 1553+113              & 1RXSJ155543.2+111114          & 238.92935                           & 11.1901014              & 0.36                         & Y    & Y   &        &       & 2.31E-11                      & 8.32E-12                     \\
5   & 3C 273.0                 & 1RXS J122906.5+020311         & 187.2779154                            & 2.0523883                  & 0.158                        & Y    & Y   & 2      & 1     & 5.19E-11                      & 8.20E-12                     \\
6   & MRK 421                  & 1RXS J110427.1+381231         & 166.1138079                         & 38.2088331              & 0.03                         & N    & Y   & 2      &       & 2.48E-10                      & 7.44E-12                     \\
7   & S5 0836+71               & 1RXS J084125.1+705342         & 130.3515217                            & 70.8950481                 & 2.172                        & N    & N   & 1.5      &       & 6.70E-12                      & 6.70E-12                     \\
8   & 1RXS J142239.1+580159    &                               & 215.662064                          & 58.032084               & 0.6349                       & Y    & Y   & 1      & 1     & 1.02E-11                      & 6.48E-12                     \\
9   & 1RXS J151747.3+652522    & 87GB 151706.4+653539          & 229.448333                          & 65.423306               & 0.702                        & Y    & Y   & 1      &       & 9.03E-12                      & 6.34E-12                     \\
10  & 1ES 1028+511             & 1RXS J103118.6+505341         & 157.8271600                            & 50.8932828                 & 0.361                        & Y    & Y   & 1      &       & 1.64E-11                      & 5.94E-12                     \\
11  & 1RXS J022716.6+020154    & RXS J02272+0201               & 36.819083                           & 2.033472                & 0.457                        & N    & N   &       &       & 1.30E-11                      & 5.92E-12                     \\
12  & 1RXS J110337.7-232931    & H 1101-232                    & 165.9067083                         & -23.492                 & 0.186                        & N    & Y   &        & 1     & 2.86E-11                      & 5.32E-12                     \\
13  & 3C 454.3                 & 1RXS J225358.0+160855         & 343.4906163                             & 16.1482111                  & 0.859                        & Y    & Y   & 1.5      &       & 5.74E-12                      & 4.93E-12                     \\
14  & 1RXS J150759.8+041511    & {[}VV2006{]} J150759.7+041512 & 226.998885                          & 4.2533289               & 1.701                        & Y    & N   & 1.5      &       & 4.80E-12                      & 4.80E-12                     \\
15  & 1RXS J122121.7+301041    & PG 1218+304                   & 185.3414208                         & 30.176975               & 0.182                        & Y    & Y   & 1.5      &       & 2.30E-11                      & 4.19E-12                     \\
16  & 3C 279                   & 1RXS J125611.2-054719         & 194.0465271 & -5.7893119               & 0.5362                       & N    & Y   &        &       & 7.72E-12                      & 4.14E-12                     \\
17  & 1RXS J003334.6-192130    & KUV 00311-1938                & 8.3925                              & -19.35925               & 0.61                         & N    & N   &        &       & 6.68E-12                      & 4.08E-12                     \\
18  & 1RXS J111706.3+201410    & RXS J11171+2014               & 169.2760833                         & 20.2353889              & 0.138                        & Y    & N   &  2      & 1     & 2.93E-11                      & 4.05E-12                     \\
19  & SHBL J012308.7+342049    & 1ES 0120+340                  & 20.785988                              & 34.346850                  & 0.272                        & Y    & Y   & 1      &       & 1.47E-11                      & 3.99E-12                     \\
20  & 1H 0414+009              & 1RXS J041652.6+010532         & 64.2187083 & 1.0899722                  & 0.287                        & Y    & Y   & 1      & 1     & 1.35E-11                      & 3.89E-12                     \\
21  & H2356-309                & 1RXS J235908.0-303740         & 359.782958                          & -30.628167              & 0.1651                       & N    & Y   & 1.5      &       & 2.32E-11                      & 3.83E-12                     \\
22  & 1RXS J050756.6+673721    & 1ES 0502+675                  & 76.984375                           & 67.6234444              & 0.314                        & Y    & Y   & 1.5      &       & 1.08E-11                      & 3.40E-12                     \\
23  & 2MASX J14283260+4240210  & 1H1426+428                    & 217.13583                           & 42.67253                & 0.129                        & Y    & Y   & 1      &       & 2.54E-11                      & 3.27E-12                     \\
24  & RBS 315                  &                               & 36.2694533                          & 18.7802128              & 2.69                         & N    & N   &        &       & 3.13E-12                      & 3.13E-12                     \\
25  & PG 1407+265              & 1RXS J140924.1+261827         & 212.349634                            & 26.305865                 & 0.947                        & Y    & Y   & 1.5      &       & 3.30E-12                      & 3.13E-12                     \\
26  & 1RXS J032540.8-164607    & RXS J03256-1646               & 51.421208                           & -16.771361              & 0.291                        & N    & N   &        &       & 1.06E-11                      & 3.10E-12                     \\
27  & PG 1437+398              & 1RXSJ143917.7+393248          & 219.8228958                         & 39.5452417              & 0.3437                       & Y    & Y   & 1      &       & 8.87E-12                      & 3.05E-12                     \\
28  & KUV 18217+6419           & 1RXS J182157.4+642051         & 275.488811                            & 64.343437                 & 0.297                        & Y    & Y   & 2      & 1     & 9.66E-12                      & 2.87E-12                     \\
29  & 1RXS J093037.1+495028    & 1ES 0927+500                  & 142.6566271                         & 49.8404308              & 0.186                        & Y    & Y   & 1      & 1     & 1.44E-11                      & 2.68E-12                     \\
30  & 1RXS J141756.8+254329    & 2E 1415+2557                  & 214.4861121                         & 25.72395                & 0.237                        & Y    & N   & 1.5      & 1     & 1.10E-11                      & 2.60E-12                     \\
31  & PG 1246+586              & 1RXSJ124818.9+582031          & 192.0782692                         & 58.34131                & 0.8474                       & Y    & Y   & 1      &       & 2.87E-12                      & 2.43E-12                     \\
32  & 1RXS J050938.3-040037    & 4U 0506-03                    & 77.409083                           & -4.012639               & 0.304                        & N    & Y   & 1.5      &       & 7.95E-12                      & 2.42E-12                     \\
33  & 1H 0419-577              & 1RXS J042601.6-571202         & 66.503018 & -57.200270                 & 0.104                        & N    & Y   &        &       & 2.25E-11                      & 2.34E-12                     \\
34  & PKS 0405-12              & 1RXSJ040748.7-121133          & 61.9517954 & -12.1935164                & 0.574                        & N    & Y   & 1      &       & 3.67E-12                      & 2.11E-12                     \\
35  & PKS 0558-504             & 1RXS J055946.5-502640         & 89.947417 & -50.447889               & 0.137                        & N    & Y   & 1.5      &       & 1.44E-11                      & 1.97E-12                     \\
36  & QSO B1959+650            & 1ES 1959+65.0          & 299.9993837                              & 65.1485144                  & 0.047                        & N    & Y   & 2       &       & 4.13E-11                      & 1.94E-12                     \\
37  & 1RXS J012338.2-231100    & RXS J01236-2311               & 20.909958                           & -23.182917              & 0.404                        & N    & N   &        &       & 4.80E-12                      & 1.94E-12                     \\
38  & 1RXS J100811.5+470526    & RXS J10081+4705               & 152.047625                          & 47.089283               & 0.343                        & Y    & N   & 1      &       & 5.65E-12                      & 1.94E-12                     \\
39  & PKS 2126-158             & {[}HB89{]} 2126-158           & 322.3007325                         & -15.6447333             & 3.268                        & N    & Y   & 1      & 1     & 1.91E-12                      & 1.91E-12                     \\
40  & QSO B0347-121            & 1RXS J034922.8-115923   & 57.346583 & -11.990833               & 0.18                         & Y    & Y   & 1      &       & 1.06E-11                      & 1.90E-12                     \\
41  & 1RXSJ055806.6-383829     &                               & 89.526958                           & -38.642139              & 0.302                        & N    & N   & 1      &       & 5.91E-12                      & 1.79E-12                     \\
42  & PKS 0548-322             & 1RXS J055040.7-321619         & 87.668981 & -32.271428                 & 0.069                        & N    & Y   & 2      & 1     & 2.38E-11                      & 1.64E-12                     \\
43  & B2 1721+34               & 1RXS J172320.5+341756         & 260.8366496 & 34.2994347                 & 0.206                        & Y    & N   & 1      & 1     & 7.75E-12                      & 1.60E-12                     \\
44  & 1RXS J121752.1+300705    & B2 1215+30                    & 184.4670079                         & 30.1168431              & 0.13                         & Y    & Y   & 1.5       &       & 1.19E-11                      & 1.55E-12                     \\
45  & 1RXS J144207.7+352632    & MARK  478                     & 220.5310975                         & 35.4397006              & 0.077                        & Y    & Y   & 2      & 1     & 1.99E-11                      & 1.54E-12                     \\
46  & 1RXS J014822.3-275828    & RXS J01483-2758               & 27.092917                           & -27.973611              & 0.121                        & N    & N   &        &       & 1.26E-11                      & 1.52E-12                     \\
47  & 1RXS J134852.6+263541    & PKS 1346+26                   & 207.22083                           & 26.59556                & 0.063                        & Y    & Y   & 1.5      & 1     & 2.40E-11                      & 1.51E-12                     \\
48  & MS0737.9+7441            &                               & 116.0219167 & 74.5660000                 & 0.315                        & N    & Y   &       & 1     & 4.76E-12                      & 1.50E-12                     \\
49  & 1RXS J063547.2-751617    & ICRF J063546.5-751616         & 98.9437829                          & -75.2713372             & 0.651                        & N    & Y   &        &       & 2.29E-12                      & 1.49E-12                     \\
50  & 1RXS J101504.3+492604    & GB 1011+496                   & 153.7672492                         & 49.4335289              & 0.2                          & Y    & Y   & 1      &       & 7.34E-12                      & 1.47E-12                     \\
51  & S5 0716+71               &                               & 110.4727017                         & 71.3434342              & 0.3                          & N    & Y   & 2      & 1     & 4.85E-12                      & 1.46E-12                     \\
52  & 1RXS J000559.1+160955    & 87GB 000325.3+155311          & 1.496817                            & 16.163615               & 0.4509                       & Y    & Y   & 1      & 1     & 3.18E-12                      & 1.43E-12                     \\
53  & PKS 2005-489             & 1RXS J200925.6-484953         & 302.3557942                         & -48.8315889             & 0.071                        & N    & Y   & 2      & 1     & 2.00E-11                      & 1.42E-12                     \\
54  & 1RXS J224520.3-465212    & RXS J22453-4652               & 341.334583                          & -46.869833              & 0.201                        & N    & Y   & 1.5      & 1     & 6.89E-12                      & 1.38E-12                     \\
55  & 2MASX J11363009+6737042  & 1RXS J113630.9+673708         & 174.125331 & 67.617887                 & 0.1342                       & Y    & N   & 1      &       & 1.03E-11                      & 1.38E-12                     \\
56  & MRK 509                  &                               & 311.0405769                         & -10.7234838             & 0.0344                       & N    & Y   &        &       & 3.81E-11                      & 1.31E-12                     \\
57  & HE 1029-1401             &                               & 157.97625                           & -14.280833              & 0.086                        & N    & Y   & 2      & 1     & 1.51E-11                      & 1.30E-12                     \\
58  & 3C 382                   &                               & 278.764125                          & 32.6963333              & 0.0579                       & N    & Y   & 1.5       &       & 2.17E-11                      & 1.26E-12                     \\
59  & 1RXS J123137.5+704417    & FBS 1229+710                  & 187.90182                           & 70.737258               & 0.208                        & N    & N   &        &       & 5.91E-12                      & 1.23E-12                     \\
60  & MS 01172-2837            &                               & 19.898333                           & -28.358611              & 0.349                        & N    & Y   &        &       & 3.40E-12                      & 1.19E-12                     \\
61  & Fairall 9                &                               & 20.940736                           & -58.805784              & 0.047                        & N    & Y   &        &       & 2.42E-11                      & 1.14E-12                     \\
62  & 1RXS J095652.4+411524    & PG 0953+415                   & 149.218302                          & 41.256181               & 0.239                        & Y    & Y   & 1.5       &       & 4.48E-12                      & 1.07E-12                     \\
63  & IRAS 13349+2438          & 1RXS J133718.8+242306         & 204.328028                            & 24.384272                 & 0.107                        & Y    & Y   & 2      & 1     & 1.00E-11                      & 1.07E-12                     \\
64  & PG 0804+761              &                               & 122.744167                          & 76.045139               & 0.1                          & N    & Y   &        &       & 1.07E-11                      & 1.07E-12                     \\
65  & 1RXS J022815.6-405712    &                               & 37.06333                            & -40.95444               & 0.495                        & N    & Y   & 1      & 1     & 2.14E-12                      & 1.06E-12                     \\
66  & TON 1388                 &                               & 169.786158                          & 21.321668               & 0.1765                       & Y    & Y   & 1      & 1     & 5.99E-12                      & 1.06E-12                     \\
67  & TON S180                 &                               & 14.333101                           & -22.383083              & 0.062                        & N    & N   &        &       & 1.66E-11                      & 1.03E-12                     \\
68  & IRAS-F22456-5125         &                               & 342.171583                          & -51.165028              & 0.1016                       & N    & Y   &        &       & 9.96E-12                      & 1.01E-12                     \\
69  & PKS 2135-14              &                               & 324.4382192                         & -14.5488358             & 0.2005                       & N    & N   & 1.5      &       & 5.04E-12                      & 1.01E-12                     \\
70  & TON 28                   &                               & 151.010894                          & 28.926496               & 0.329                        & Y    & Y   & 1      &       & 2.93E-12                      & 9.63E-13                     \\
71  & PKS 1830-210             & QSO B1830-210                 & 278.4162007                         & -21.0610479             & 2.507                        & N    & Y   &        &       & 9.15E-13                      & 9.15E-13                     \\
72  & AKN 564                  &                               & 340.6639388                         & 29.7253637              & 0.0247                       & Y    & Y   &        &       & 3.67E-11                      & 9.05E-13                     \\
73  & Q 1252+0200              &                               & 193.83207                           & 1.736726                & 0.345                        & Y    & Y   & 1      &       & 2.59E-12                      & 8.94E-13                     \\
74  & PG 0052+251              &                               & 13.717167                           & 25.427500                 & 0.1545                       & Y    & Y   & 1      & 1     & 5.37E-12                      & 8.29E-13                     \\
75  & MRK 180                  &                               & 174.110035                          & 70.157585               & 0.046                        & N    & Y   &        &       & 1.79E-11                      & 8.25E-13                     \\
76  & PG 1100+772              &                               & 166.0570288                         & 76.9827847              & 0.312                        & Y    & Y   & 2      &       & 2.50E-12                      & 7.81E-13                     \\
77  & 3C 48                    & 1RXS J013741.7+330931         & 24.4220808                          & 33.1597594              & 0.367                        & Y    & Y   & 1      &       & 2.12E-12                      & 7.80E-13                     \\
78  & TON S210                 & 1RXS J012151.5-282048         & 20.464585                             & -28.349281                & 0.117                        & N    & Y   &        &       & 6.53E-12                      & 7.63E-13                     \\
79  & AKN 120                  &                               & 79.047588                           & -0.1498268              & 0.0327                       & Y    & Y   &       & 1     & 2.33E-11                      & 7.62E-13                     \\
80  & ESO 141-G55              & 1RXS J192115.3-584011         & 290.308902                          & -58.670307              & 0.037                        & N    & Y   &        &       & 2.05E-11                      & 7.58E-13                     \\
81  & PG 1613+658              &                               & 243.4882486                         & 65.7193264              & 0.129                        & N    & Y   & 2      & 1     & 5.58E-12                      & 7.20E-13                     \\
82  & MARK 813                 &                               & 216.8544292                         & 19.8309823              & 0.1105                       & Y    & Y   & 1      &       & 6.38E-12                      & 7.05E-13                     \\
83  & ESO 198-024              &                               & 39.581889                           & -52.192377              & 0.0455                       & N    & N   &        &       & 1.48E-11                      & 6.74E-13                     \\
84  & NAB 0205+02              & Mrk 586                       & 31.957767                           & 2.715447                & 0.1555                       & Y    & Y   & 1      &       & 4.12E-12                      & 6.41E-13                     \\
85  & MRK 279                  &                               & 208.2643618                         & 69.3082128              & 0.0304                       & N    & Y   & 2      & 1     & 2.10E-11                      & 6.38E-13                     \\
86  & 1RXS J040501.1-371110    & ESO 359-G19                   & 61.256571                           & -37.187524              & 0.056                        & N    & N   &        &       & 1.14E-11                      & 6.37E-13                     \\
87  & 3C 390.3                 &                               & 280.5374579                         & 79.7714242              & 0.0561                       & N    & Y   & 2      & 1     & 1.13E-11                      & 6.35E-13                     \\
88  & NGC 5548                 &                               & 214.4980583                         & 25.13679                & 0.0171                       & Y    & Y   & 2      & 1     & 3.48E-11                      & 5.95E-13                     \\
89  & MRK 205                  &                               & 185.434254                          & 75.310787               & 0.07                         & N    & Y   & 1      &       & 7.71E-12                      & 5.40E-13                     \\
90  & PHL 1092                 &                               & 24.982294                           & 6.322921                & 0.396                        & Y    & Y   & 1      &       & 1.35E-12                      & 5.34E-13                     \\
91  & SBS 1419+480             &                               & 215.373953                          & 47.790149               & 0.0723                       & Y    & N   & 1      & 1     & 5.48E-12                      & 3.96E-13                     \\
92  & 1E 0514-0030             & 1RXS J051633.3-002617         & 79.139167                           & -0.453611               & 0.292                        & Y    & Y   & 1      &       & 1.30E-12                      & 3.79E-13                     \\
93  & ESO 511-G030             & 1RXS J141922.5-263842         & 214.843417                          & -26.644722              & 0.022                        & N    & N   & 2      & 1     & 1.35E-11                      & 2.98E-13                     \\
94  & CTS G03.04               & 1RXS J193804.4-510950         & 294.518292                          & -51.163778              & 0.04                         & N    & N   &        &       & 6.91E-12                      & 2.77E-13                     \\
95  & RXS J17414+0348          & 1RXS J174128.1+034848         & 265.36775                           & 3.814694                & 0.03                         & N    & N   & 2      & 1     & 9.16E-12                      & 2.75E-13                     \\
96  & NGC 3516                 &                               & 166.6978756                         & 72.5685767              & 0.0088                       & N    & Y   &        &       & 2.54E-11                      & 2.24E-13                     \\
97  & H 1934-063               & 1RXS J193732.8-061305         & 294.387542                          & -6.218000                  & 0.01                         & N    & N   &        &       & 2.01E-11                      & 2.01E-13                     \\
98  & KUV 22497+1439           &                               & 343.033587                          & 14.913717               & 0.13                         & Y    & N   & 1      & 1     & 1.47E-12                      & 1.91E-13                     \\
99  & MCG-6-30-15              & ESO 383- G 035                & 203.974083                          & -34.295583              & 0.0077                       & N    & Y   &        &       & 2.40E-11                      & 1.85E-13                     \\
100 & H 1722+119               & 1RXS J172504.4+115218         & 261.2680867                         & 11.8709633              & 0.018                        & N    & Y   &        &       & 9.55E-12                      & 1.72E-13                     \\
101 & NGC 3783                 &                               & 174.757167                          & -37.738583              & 0.0097                       & N    & Y   &        &       & 1.48E-11                      & 1.44E-13                     \\
102 & NGC 1068                 &                               & 40.6696292                          & -0.0132806              & 0.0038                       & Y    & Y   &        &       & 1.36E-11                      & 5.16E-14                     \\
103 & NGC 4051                 &                               & 180.7900601                         & 44.5313344              & 0.0023                       & N    & Y   & 1.5    &       & 1.70E-11                      & 3.90E-14                     \\
104 & NGC 4151                 &                               & 182.6357449                         & 39.4057299              & 0.0043                       & Y    & Y   & 1.5    &       & 4.84E-12                      & 2.08E-14                     \\

\enddata
\end{deluxetable} 

\begin{deluxetable}{clrrrrrr}
\tabletypesize{\scriptsize}
\tablecaption{Observing Strategy and Number of Absorption Systems}
\label{tab:observing_strategy}
\tablewidth{0pt}
\tablehead{\colhead{\#} & \colhead{Name} & \colhead{z} & \colhead{$t_{exp}$} & \colhead{$t_{tot}$} & \colhead{$\Sigma z$} & \colhead{No. Abs} & \colhead{Sum Abs}}
\startdata
1	&	PKS 2155-304	&	0.1160	&	0.10	&	0.10	&	0.1160	&	0.7772	&	0.78	   \\
2	&	Ton 116	&	1.0654	&	0.50	&	0.60	&	1.1160	&	6.7000	&	7.48	   \\
3	&	1RXS J153501.1+532042	&	0.8900	&	0.47	&	1.06	&	2.0060	&	5.9630	&	13.44	   \\
4	&	PG 1553+113 	&	0.3600	&	0.19	&	1.26	&	2.3660	&	2.4120	&	15.85	   \\
5	&	3C 273.0	&	0.1580	&	0.10	&	1.36	&	2.5240	&	1.0586	&	16.91	   \\
6	&	MRK 421	&	0.0300	&	0.10	&	1.46	&	2.5540	&	0.2011	&	17.11	   \\
7	&	S5 0836+71	&	2.1720	&	0.67	&	2.12	&	3.5540	&	6.7000	&	23.81	   \\
8	&	1RXS J142239.1+580159	&	0.6349	&	0.44	&	2.56	&	4.1889	&	4.2540	&	28.07	   \\
9	&	1RXS J151747.3+652522	&	0.7020	&	0.49	&	3.06	&	4.8909	&	4.7034	&	32.77	   \\
10	&	1ES 1028+511	&	0.3610	&	0.27	&	3.33	&	5.2519	&	2.4187	&	35.19	   \\
11	&	1RXS J022716.6+020154	&	0.4570	&	0.34	&	3.67	&	5.7089	&	3.0619	&	38.25	   \\
12	&	1RXS J110337.7-232931	&	0.1860	&	0.16	&	3.83	&	5.8949	&	1.2462	&	39.50	   \\
13	&	3C 454.3	&	0.8590	&	0.78	&	4.61	&	6.7539	&	5.7553	&	45.25	   \\
14	&	1RXS J150759.8+041511	&	1.7010	&	0.93	&	5.54	&	7.7539	&	6.7000	&	51.95	   \\
15	&	1RXS J122121.7+301041	&	0.1820	&	0.19	&	5.73	&	7.9359	&	1.2194	&	53.17	   \\
16	&	3C 279	&	0.5362	&	0.58	&	6.31	&	8.4721	&	3.5925	&	56.76	   \\
17	&	1RXS J003334.6-192130	&	0.6100	&	0.67	&	6.98	&	9.0821	&	4.0870	&	60.85	   \\
18	&	1RXS J111706.3+201410	&	0.1380	&	0.15	&	7.13	&	9.2201	&	0.9246	&	61.77	   \\
19	&	SHBL J012308.7+342049	&	0.2720	&	0.30	&	7.43	&	9.4921	&	1.8224	&	63.60	   \\
20	&	1H 0414+009	&	0.2870	&	0.33	&	7.76	&	9.7791	&	1.9229	&	65.52	   \\
21	&	H2356-309	&	0.1651	&	0.19	&	7.95	&	9.9442	&	1.1061	&	66.63	   \\
22	&	1RXS J050756.6+673721	&	0.3140	&	0.41	&	8.37	&	10.2582	&	2.1038	&	68.73	   \\
23	&	2MASX J14283260+4240210	&	0.1290	&	0.18	&	8.54	&	10.3872	&	0.8643	&	69.59	   \\
24	&	RBS 315	&	2.6900	&	1.42	&	9.97	&	11.3872	&	6.7000	&	76.29	   \\
25	&	PG 1407+265	&	0.9470	&	1.35	&	11.32	&	12.3342	&	6.3449	&	82.64	   \\
26	&	1RXS J032540.8-164607	&	0.2910	&	0.42	&	11.74	&	12.6252	&	1.9497	&	84.59	   \\
27	&	PG 1437+398 	&	0.3437	&	0.50	&	12.24	&	12.9689	&	2.3025	&	86.89	   \\
28	&	KUV 18217+6419	&	0.2970	&	0.46	&	12.70	&	13.2659	&	1.9899	&	88.88	   \\
29	&	1RXS J093037.1+495028	&	0.1860	&	0.31	&	13.01	&	13.4519	&	1.2462	&	90.13	   \\
30	&	1RXS J141756.8+254329	&	0.2370	&	0.41	&	13.42	&	13.6889	&	1.5879	&	91.72	   \\
31	&	PG 1246+586 	&	0.8474	&	1.55	&	14.97	&	14.5363	&	5.6775	&	97.39	   \\
32	&	1RXS J050938.3-040037	&	0.3040	&	0.56	&	15.54	&	14.8403	&	2.0368	&	99.43	   \\
33	&	1H 0419-577	&	0.1040	&	0.20	&	15.73	&	14.9443	&	0.6968	&	100.13	   \\
34	&	PKS 0405-12	&	0.5740	&	1.22	&	16.95	&	15.5183	&	3.8458	&	103.97	   \\
35	&	PKS 0558-504	&	0.1370	&	0.31	&	17.26	&	15.6553	&	0.9179	&	104.89	   \\
36	&	QSO B1959+650	&	0.0470	&	0.11	&	17.37	&	15.7023	&	0.3149	&	105.21	   \\
37	&	1RXS J012338.2-231100	&	0.4040	&	0.93	&	18.30	&	16.1063	&	2.7068	&	107.91	   \\
38	&	1RXS J100811.5+470526	&	0.3430	&	0.79	&	19.09	&	16.4493	&	2.2981	&	110.21	   \\
39	&	PKS 2126-158	&	3.2680	&	2.34	&	21.43	&	17.4493	&	6.7000	&	116.91	   \\
40	&	QSO B0347-121	&	0.1800	&	0.42	&	21.85	&	17.6293	&	1.2060	&	118.12	   \\
41	&	1RXSJ055806.6-383829	&	0.3020	&	0.76	&	22.61	&	17.9313	&	2.0234	&	120.14	   \\
42	&	PKS 0548-322	&	0.0690	&	0.19	&	22.80	&	18.0003	&	0.4623	&	120.60	   \\
43	&	B2 1721+34	&	0.2060	&	0.58	&	23.37	&	18.2063	&	1.3802	&	121.98	   \\
44	&	1RXS J121752.1+300705	&	0.1300	&	0.37	&	23.75	&	18.3363	&	0.8710	&	122.85	   \\
45	&	1RXS J144207.7+352632	&	0.0770	&	0.22	&	23.97	&	18.4133	&	0.5159	&	123.37	   \\
46	&	1RXS J014822.3-275828	&	0.1210	&	0.36	&	24.33	&	18.5343	&	0.8107	&	124.18	   \\
47	&	1RXS J134852.6+263541	&	0.0630	&	0.19	&	24.51	&	18.5973	&	0.4221	&	124.60	   \\
48	&	MS0737.9+7441	&	0.3150	&	0.94	&	25.45	&	18.9123	&	2.1105	&	126.71	   \\
49	&	1RXS J063547.2-751617	&	0.6510	&	1.95	&	27.39	&	19.5633	&	4.3617	&	131.07	   \\
50	&	1RXS J101504.3+492604	&	0.2000	&	0.61	&	28.00	&	19.7633	&	1.3400	&	132.41	   \\
51	&	S5 0716+71	&	0.3000	&	0.92	&	28.92	&	20.0633	&	2.0100	&	134.42	   \\
52	&	1RXS J000559.1+160955	&	0.4509	&	1.40	&	30.33	&	20.5142	&	3.0210	&	137.45	   \\
53	&	PKSÿ2005-489	&	0.0710	&	0.22	&	30.55	&	20.5852	&	0.4757	&	137.92	   \\
54	&	1RXS J224520.3-465212	&	0.2010	&	0.65	&	31.20	&	20.7862	&	1.3467	&	139.27	   \\
55	&	2MASX J11363009+6737042	&	0.1342	&	0.43	&	31.63	&	20.9204	&	0.8991	&	140.17	   \\
56	&	MRK 509 	&	0.0344	&	0.12	&	31.75	&	20.9548	&	0.2305	&	140.40	   \\
57	&	HE 1029-1401	&	0.0860	&	0.29	&	32.04	&	21.0408	&	0.5762	&	140.97	   \\
58	&	3C 382	&	0.0579	&	0.21	&	32.25	&	21.0987	&	0.3877	&	141.36	   \\
59	&	1RXS J123137.5+704417	&	0.2080	&	0.76	&	33.00	&	21.3067	&	1.3936	&	142.75	   \\
60	&	MS 01172-2837	&	0.3490	&	1.31	&	34.32	&	21.6557	&	2.3383	&	145.09	   \\
61	&	Fairall 9	&	0.0470	&	0.18	&	34.50	&	21.7027	&	0.3150	&	145.41	   \\
62	&	1RXS J095652.4+411524	&	0.2390	&	1.00	&	35.50	&	21.9417	&	1.6013	&	147.01	   \\
63	&	IRAS 13349+2438	&	0.1070	&	0.45	&	35.94	&	22.0487	&	0.7169	&	147.73	   \\
64	&	PG 0804+761	&	0.1000	&	0.42	&	36.36	&	22.1487	&	0.6700	&	148.40	   \\
65	&	1RXS J022815.6-405712 	&	0.4950	&	2.08	&	38.45	&	22.6437	&	3.3165	&	151.71	   \\
66	&	TON 1388	&	0.1765	&	0.74	&	39.19	&	22.8202	&	1.1826	&	152.90	   \\
67	&	TON S180	&	0.0620	&	0.27	&	39.46	&	22.8822	&	0.4154	&	153.31	   \\
68	&	IRAS-F22456-5125	&	0.1016	&	0.45	&	39.91	&	22.9838	&	0.6807	&	153.99	   \\
69	&	PKS 2135-14	&	0.2005	&	0.89	&	40.80	&	23.1842	&	1.3431	&	155.33	   \\
70	&	TON 28	&	0.3290	&	1.53	&	42.32	&	23.5132	&	2.2043	&	157.54	   \\
71	&	PKS 1830-210	&	2.5070	&	4.88	&	47.20	&	24.5132	&	6.7000	&	164.24	   \\
72	&	AKN 564	&	0.0247	&	0.12	&	47.32	&	24.5379	&	0.1654	&	164.40	   \\
73	&	Q 1252+0200	&	0.3450	&	1.72	&	49.04	&	24.8829	&	2.3115	&	166.72	   \\
74	&	PG 0052+251	&	0.1545	&	0.83	&	49.87	&	25.0374	&	1.0348	&	167.75	   \\
75	&	MRK 180	&	0.0460	&	0.25	&	50.12	&	25.0834	&	0.3082	&	168.06	   \\
76	&	PG 1100+772	&	0.3120	&	1.78	&	51.91	&	25.3954	&	2.0904	&	170.15	   \\
77	&	3C 48	&	0.3670	&	2.10	&	54.01	&	25.7624	&	2.4589	&	172.61	   \\
78	&	TON S210	&	0.1170	&	0.68	&	54.69	&	25.8794	&	0.7839	&	173.39	   \\
79	&	AKN 120	&	0.0327	&	0.19	&	54.88	&	25.9121	&	0.2192	&	173.61	   \\
80	&	ESOÿ141-G55	&	0.0370	&	0.22	&	55.10	&	25.9491	&	0.2479	&	173.86	   \\
81	&	PG 1613+658	&	0.1290	&	0.80	&	55.90	&	26.0781	&	0.8643	&	174.72	   \\
82	&	MARK 813	&	0.1105	&	0.70	&	56.60	&	26.1886	&	0.7404	&	175.46	   \\
83	&	ESO 198-024	&	0.0455	&	0.30	&	56.90	&	26.2341	&	0.3049	&	175.77	   \\
84	&	NAB 0205+02	&	0.1555	&	1.08	&	57.99	&	26.3896	&	1.0421	&	176.81	   \\
85	&	MRK 279	&	0.0304	&	0.21	&	58.20	&	26.4200	&	0.2037	&	177.01	   \\
86	&	1RXS J040501.1-371110	&	0.0560	&	0.39	&	58.59	&	26.4760	&	0.3752	&	177.39	   \\
87	&	3C 390.3	&	0.0561	&	0.39	&	58.98	&	26.5321	&	0.3759	&	177.77	   \\
88	&	NGC 5548	&	0.0171	&	0.13	&	59.11	&	26.5492	&	0.1146	&	177.88	   \\
89	&	MRK 205	&	0.0700	&	0.58	&	59.69	&	26.6192	&	0.4690	&	178.35	   \\
90	&	PHL 1092	&	0.3960	&	3.31	&	63.00	&	27.0152	&	2.6532	&	181.00	   \\
91	&	SBS 1419+480	&	0.0723	&	0.81	&	63.82	&	27.0875	&	0.4844	&	181.49	   \\
92	&	1Eÿ0514-0030	&	0.2920	&	3.44	&	67.25	&	27.3795	&	1.9564	&	183.44	   \\
93	&	ESOÿ511-G030	&	0.0220	&	0.33	&	67.58	&	27.4015	&	0.1474	&	183.59	   \\
94	&	CTSÿG03.04	&	0.0400	&	0.65	&	68.23	&	27.4415	&	0.2680	&	183.86	   \\
95	&	RXSÿJ17414+0348	&	0.0300	&	0.49	&	68.72	&	27.4715	&	0.2010	&	184.06	   \\
96	&	NGC 3516	&	0.0088	&	0.18	&	68.89	&	27.4803	&	0.0590	&	184.12	   \\
97	&	Hÿ1934-063	&	0.0100	&	0.22	&	69.12	&	27.4903	&	0.0670	&	184.19	   \\
98	&	KUV 22497+1439	&	0.1300	&	3.04	&	72.16	&	27.6203	&	0.8710	&	185.06	   \\
99	&	MCG-6-30-15	&	0.0077	&	0.19	&	72.34	&	27.6280	&	0.0516	&	185.11	   \\
100	&	Hÿ1722+119	&	0.0180	&	0.47	&	72.81	&	27.6460	&	0.1206	&	185.23	   \\
101	&	NGC 3783	&	0.0097	&	0.30	&	73.11	&	27.6557	&	0.0650	&	185.29	   \\
102	&	NGC 1068	&	0.0038	&	0.33	&	73.44	&	28.6595	&	0.0255	&	185.32	   \\
103	&	NGC 4051	&	0.0023	&	0.26	&	73.71	&	28.6618	&	0.0154	&	185.33	   \\
104	&	NGC 4151	&	0.0043	&	0.92	&	74.63	&	28.6661	&	0.0288	&	185.36	   \\
\enddata
\end{deluxetable}

\begin{deluxetable}{lccc}
\tabletypesize{\scriptsize}
\tablecaption{Mission Capabilities and Science Outcomes}
\label{tab:summary}
\tablewidth{0pt}
\tablehead{
 & & \colhead{Nominal Mission} & \colhead{Minimum Mission}\\
}
\startdata
$A_{\text{eff}}$          & (cm$^{2}$)  & 1000 & 300  \\
$R$                       &             & 3000 & 2000 \\
$5\sigma$ $W_{\text{eq}}$ & (m\AA)      & 3    & 5    \\
\hline  
\multicolumn{4}{c}{$dN/dz$} \\
\hline \\
Num. Sightlines           &             & 100  & 33   \\
Survey Time               & ($10^6$\,s) & 8    & 8    \\
Normalization Accuracy    &             & 10\% & 17\% \\
$dN/dz$ Slope             &             & full shape & poor cutoff constraint \\
Other                     &   &  some z evolution & \\
\hline 
\multicolumn{4}{c}{Halo Mass and Distribution} \\
\hline  \\
$N_{\text{AGN}}$ (External Gal.) &      & 20   & 10   \\
$N_{\text{AGN}}$ (Milky Way)     &      & 35   & 35   \\
$N_{\text{AGN}}$ (M31)           &      &  6$^a$   &  3    \\
Distribution Accuracy            &      & $\beta$ to 3\% & $\beta$ to 5\% \\
MW Optical Depth Study           &      & Yes  & Yes  \\
\enddata
\tablecomments{The increased effective area for the Nominal Mission enables most of the science
goals outlined in this paper. The Minimum Mission would still make substantial progress,
but would not be useful for the majority of sight lines. 
\newline
Footnote a:  An additional 1 Msec of observing time increases number to 11 targets.}
\end{deluxetable}

\newpage


\end{spacing}
\end{document}